\documentclass{aa}  

\usepackage{graphicx}
\usepackage{txfonts}
\usepackage{hyperref}
\def\msun{{\rm M}_{\odot}} 
 
\def\rvir{R_{\rm vir}}
\def\tff{t_{\rm ff}}
\def\tcool{t_{\rm cool}}

\def\kms{ {\rm km}\,{\rm s}^{-1} }
\def\cc{{\rm cm}^{-3}}
\def\yr{{\rm yr}}
\def\cm2{{\rm cm}^{-2}}

\newcommand{\gasoline}{\textsc{Gasoline}}
\newcommand{\cloudy}{\textsc{Cloudy}}

\begin{document}

   \title{The origin of cold gas in the circumgalactic medium}

   % \subtitle{subtitle.. }

   \author{Davide Decataldo
          \inst{1},
          %\and
          Sijing Shen\inst{1}, 
          Lucio Mayer\inst{2},
          Bernhard Baumschlager\inst{1},
          Piero Madau\inst{3}
          }

   \institute{Institute of Theoretical Astrophysics, University of Oslo, P.O. Box 1029, Blindern, 0315 Oslo, Norway\\
              \email{davide.decataldo@astro.uio.no}
        \and
             Center for Theoretical Astrophysics and Cosmology, Institute for Computational Science, University of Zurich, \\Winterthurerstrasse 190, 8057 Zürich, Switzerland
        \and 
            Department of Astronomy and Astrophysics, University of California, 1156 High Street, Santa Cruz, CA 95064, USA
             }

   \date{Received xxx; accepted xxx}

\titlerunning{Origin of cold gas in the CGM}
\authorrunning{Decataldo D., Shen S., Mayer L., Baumschlager B., Madau P.}

% \abstract{}{}{}{}{} 
% 5 {} token are mandatory
 
  \abstract
   {
    The presence of cold gas ($T \lesssim 10^4$ K)  in the circumgalactic medium (CGM) of galaxies has been confirmed in observations and in high-resolution simulations, but its origin is still   a puzzle. Possible mechanisms are cold accretion from the intergalactic medium (IGM), clumps embedded in outflows and transported from the disk, and gas detaching from the hot CGM phase via thermal instabilities.
   }
   {
   In this work we   characterize the history of cold CGM gas in order to identify the dominant origin channels at different evolutionary stages of the main galaxy. 
 }
   {
   To this end, we tracked gas particles in different snapshots of the smoothed particle hydrodynamics (SPH) cosmological zoom-in simulation Eris2k.  We performed a backward tracking of cold gas, starting from different redshifts until we could identify one of the followings origins for the particle: cold inflow,  ejection from the disk, cooling down in situ, or stripping from a satellite. We also performed a forward tracking of gas in different components of the galaxy (such as the disk and outflows).
   }
   {
   We find a clear transition between two epochs. For $z>2$, most cold gas (up to 80\%) in the CGM comes from cold accretion streams as the galaxy is accreting in the  cold mode  from the IGM. At lower $z$, gas either cools down in situ  after several recycles (with 10-20\% of the gas cooling in outflow), or it is  ejected directly from the disk (up to 30\%). Outflows have a major contribution to the cold CGM gas budget at $z<1$, with almost 50\% of the hot gas cooling in outflow. Finally, we discuss possible mechanisms for CGM cooling, showing that the thermally unstable gas  with $t_{\rm cool}/t_{\rm ff}<1$ (precipitation-regulated feedback) is abundant up to $r\sim 100$ kpc and cooling times are shorter than 50 Myr for densities $n>10^{-2}\,\cc$.

   }
   
 %  }


   \maketitle

%-------------------------------------------------------------------
%-------------------------------------------------------------------
%-------------------------------------------------------------------
%
%
%
%
%-------------------------------------------------------------------
%-------------------------------------------------------------------
\section{Introduction}

The circumgalactic medium (CGM) of a galaxy includes all the gas in the potential well of the dark matter halo that extends outside the disk and up to a few times the virial radius. There is evidence that this gas has a complex multi-phase structure, including hot virialized gas at $T\sim 10^{6-7}$ K \citep{allen2008improved, dai2010baryon} and cold gas with hydrogen in atomic and molecular form \citep[$T<10^5$ K, ][]{prochaska2013substantial, battaia2016stacked, meiksin2017gas, fossati2021muse}.

This multi-phase structure has been observed mainly via absorption-line spectroscopy. The hot phase is traced by atoms with high ionization potential, such as NV, OVI, and NeVIII  \citep{stocke2006galaxy, tumlinson2011large, savage2011cos, meiring2013qso, pachat2017detection, burchett2018warm, manuwal2021cos}, while atoms with low ionization potential are excited at temperatures $\sim 10^4$ K, such as MgII and SiII \citep{steidel2010structure, gauthier2010incidence, matejek2012survey, prochaska2014quasars}. Using this technique, the COS-Halos survey \citep{tumlinson2013cos} showed abundant cold gas in halos with stellar mass in the range $10^{9.6}-10^{11.3}\,\msun$, both in actively  star-forming galaxies and quenched passive galaxies. More recently, cold gas has also been detected in emission, using different tracers, such as Ly$\alpha$ \citep{arrigoni2019qso, battaia2022multiwavelength, bacon2021muse}, OII, MgII \citep{rupke2019100, burchett2021circumgalactic}, and CO(3-2) \citep{cicone2021super}. 

The presence of cold gas when the virial temperature exceeds $10^6$ K is currently a puzzle. The cooling times of low-density gas are generally very long, and hence mechanisms to locally increase the density are needed. In principle, gas could detach from the hot phase because of density perturbations triggering thermal instability \citep{field1965thermal, mccourt2012thermal}. Density can also increase due to convective instability \citep{balbus1989theory, malagoli1990numerical}, wind-driven shocks \citep{zubovas2014galaxy, ferrara2016formation, richings2018origin}, or a condensation-precipitation mechanism \citep{voit2015precipitation, voit2017global}. Another possible scenario is that cold gas is lifted directly from the disk and entrained into an outflow, which could be an explanation for the amount of outflowing molecular gas in active galactic nuclei  \citep{cicone2014massive}. This gas is subject to strong shocks due to the extremely turbulent environment and surrounded by a high-velocity hot medium, so clouds can be disrupted because of thermal conduction and Kelvin–Helmholtz instabilities \citep{mckee1990steady, ferrara2016formation}. Furthermore, the clouds are exposed to the intense ultraviolet radiation field from the center of the galaxy that is able to dissociate molecular gas, making it challenging for clouds to survive and travel up to distances of a few kiloparsecs \citep{decataldo2017molecular, decataldo2019photoevaporation}. Finally, cold gas can be found inside accretion filaments that fuel the galaxy disk. Observations \citep{rubin2010galaxies, rubin2015dissecting, lehner2014galactic} and simulations \citep{fardal2001cooling, kerevs2005galaxies, van2012cold, nelson2013moving} agree that a large fraction of gas accretes into galaxies via a ``cold mode,'' especially in smaller galaxies where accretion shocks might not be able to heat up the rapidly cooling gas in dense filaments. 

Numerical simulations are a powerful tool for improving our understanding of the origin of cold gas. Simulations can be done in two ways. First, it is possible to study the gas on small scales to understand the evolution of thermal instabilities and the dynamics of cold gas in different environments, for example  simulations of stratified media \citep{joung2012gas}, filamentary streams \citep{mandelker2019instability}, and the cloud-crushing problem \citep{scannapieco2015launching, schneider2018production, gronke2018growth}. The second way is by using  cosmological simulations, which allow us to visualize the large-scale structure, showing where the cold gas is more abundant and where it comes from; one approach is to track the gas trajectory, and hence determining if the CGM gas cools down in situ or if it is carried already in cold phase within accretion streams or outflows. By analyzing simulations of large cosmological boxes \citep{oppenheimer2018multiphase, hafen2019origins, nelson2020resolving} it is possible to obtain  statistics over many halos of different masses. On the other hand, zoom-in simulations allow us to reach a greater level of detail on a single halo, with the inclusion of more detailed physics \citep{peeples2019figuring, suresh2019zooming}.

Nevertheless, resolving the cold gas in the CGM is a challenging task, even for the most modern simulations. In particular, the cold gas mass, and the abundance of low ionization potential ions as a consequence, are typically under-reproduced in comparison to the observational results \citep[e.g., the COS-Halos survey;][]{tumlinson2011large}. It has been proposed \citep{hummels2019impact} that this is due to the lack of resolution since cold gas is usually organized in dense clumps and clouds that cannot be resolved \citep{rauch1999small, crighton2015metal, stern2016universal, rubin2018galaxies}, ending up mixed with the hot phase. Furthermore, since Lagrangian methods and common refinement schemes in Eulerian methods focus on resolving the dense gas, the diffuse phase of the CGM is generally poorly resolved. Some simulations manage to find a compromise between the improved resolution and the increase in  computational cost, adopting refinement criteria that explicitly target  the CGM, for example additional spatial refinement up to the virial radius
\citep{hummels2019impact, van2019cosmological, peeples2019figuring} and super-Lagrangian zoom schemes \citep{suresh2019zooming}. 

Most of past theoretical works discussed the origin of the CGM gas, without focusing on the cold phase. For example, \citet{ford2014tracing} track particles backward and forward from a simulation snapshot at $z=0.25$ to distinguish between gas in the process of accretion, gas ejected at a previous time, and ambient gas that always belonged to the CGM. They find ambient gas to be the dominant component (up to 85\%), with outflows becoming more important (up to 43\%) for low-mass halos ($<10^{11.5}\,\msun$). A fraction of outflowing gas is later re-accreted, and this cycle can happen multiple times \citep[gas recycling;][]{oppenheimer2010feedback, angles2017cosmic}. In a similar analysis over a set of FIRE-2 simulations, \citet{hafen2019origins} classify 60-80\% of gas as intergalactic medium (IGM) accretion and $\sim$ 10-30\% from winds, but the IGM accretion category includes any gas particles not belonging to other galaxies, hence also including  the  ambient gas in \citet{ford2014tracing}. An analysis focused on cold gas was presented in \citet{suresh2019zooming}, where galactic fountains were identified as the major source, with most gas recycled more than once. Interestingly, a comparison between two simulations with and without winds shows that winds are able to trigger gas cooling, increasing by $\sim$ 50\% the HI covering fraction. Finally, \citet{nelson2020resolving} identify cold gas clouds in the TNG50 cosmological box \citep{pillepich2019first} at  redshift $z\sim 5$, and follow their assembling history. They find that clouds form by accretion of gas into dense pre-existing seeds, which mostly originate from density perturbations induced by mergers with infalling satellite galaxies.

In this work we performed an analysis on the origin and fate of the cold CGM in gas in the cosmological zoom-in simulation Eris2k. Differently from most previous works that typically track the cold gas back in time until some location criteria are met, we consider simultaneously the position and temperature evolution of gas particles in order to understand when and where gas cooling happens. In particular, we distinguish between gas cooling down directly in the CGM (while in outflow,   in inflow, or static) and gas entering the CGM already in a cold state (from the IGM, from the disk of the main galaxy, or from a satellite). This allows us to investigate   the mechanisms responsible for gas cooling, and   the survival of cold gas clouds transported into the CGM. 

The paper is organized as follows. In Sect. 2 we present the simulation Eris2k, and describe our algorithm to track smoothed particle hydrodynamic (SPH) particles. In Sect. 3 we detail the properties of cold gas in Eris2k (Sect. 3.1), we discuss its origin (Sect. 3.2), and we analyze in more detail the gas temperature evolution by forward tracking gas in the disk (Sect. 3.3) and in outflow (Sect. 3.4). A discussion  of the cooling physics and cooling times is presented in Sect. 3.5. We  summarize our conclusions in Sect. 4.

%-------------------------------------------------------------------
%--------------------------------------------------------------------
\section{Methods}

%--------------------------------------------------------------------
\subsection{The Eris2k simulation}

\begin{figure}
\centering
\includegraphics[width=0.45\textwidth]{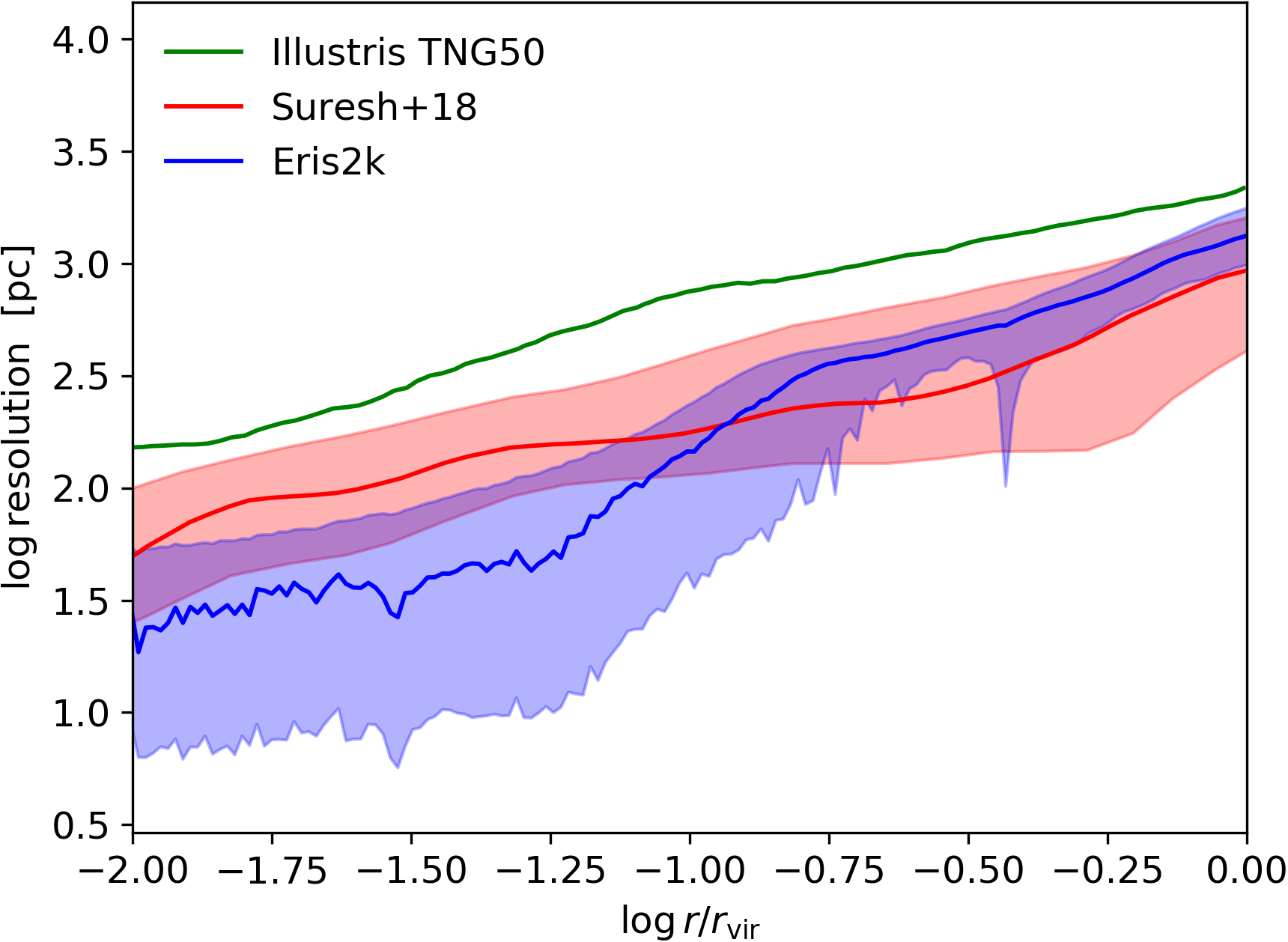}
\caption{
Gas resolution (defined as the smoothing length of SPH particles) in the simulation Eris2k as a function of   distance from the main halo center. The blue curve represents the median value in the radial bin, while the shaded area includes the variation from the 10th to the 90th percentile. The other two lines show the median resolution for a halo in Illustris TNG50 ($M_{\rm halo}\simeq 10^{12.5}\,\msun$) and the high-resolution simulation in \citet{suresh2019zooming} (with 10th to 90th percentile variation).
}
\label{smoothing_length}
\end{figure}

In this work we analyze the cosmological zoom-in simulations of a Milky
Way-like spiral galaxy, Eris2k. The simulation was run with the SPH code \gasoline \citep{wadsley2004gasoline}, assuming   flat WMAP-3 cosmology ($\Omega_M=0.24$, $\Omega_b=0.042$, $H_0=73\,{\rm km\,s^{-1}\,Mpc^{-1}}$, $n=0.96$, $\sigma_8=0.76$). The mass resolution is $m_{\rm dm}=9.8\times 10^4\,\msun$ for dark matter and $m_{\rm g}=2\times 10^4\,\msun$ for gas particles.
The ionization fractions of H and He were computed taking into account photoionization by a uniform UV background \citep{haardt2012radiative}, recombination, and collisional ionization, with rates from \citet{abel1996modeling}. Metal-line cooling rates were pre-computed with the code \cloudy~\citep{ferland1998cloudy} on the assumption of ionization equilibrium \citep{shen2010enrichment}. Star particles, representing a population of stars with a \citet{kroupa2001variation} initial mass function (IMF), are formed stochastically by converting dense ($n>100\,\cc$) and cold ($T<10^4\,{\rm K}$) gas particles. Supernova feedback is included by injecting thermal energy (with the delayed cooling approximation) and metals into the gas.

At $z\simeq 0.3$, the main halo in Erisk2k (total mass $M=7.2\times 10^{11}\,\msun$, virial radius $R_{\rm vir}=206$ kpc) hosts a Milky Way-like galaxy with  $4.9\times 10^{10}\,\msun$ in gas and $3.9\times 10^{10}\,\msun$ in stars. The galaxy has a quiet merging history, with no major mergers (ratio $>1:10$) happening for $z\lesssim 3$. The star formation rate is quite constant ($\sim 10\, \msun\yr^{-1}$) from $z\simeq 5$ to $z\simeq 2$, and then drops to a value of about $2\,\msun\yr^{-1}$ at $z\simeq 0.3$.

In Fig. \ref{smoothing_length} we show the spatial resolution (i.e., the smoothing length of the particles) in Eris2k, compared to other simulations that have been adopted in other CGM studies. The median spatial resolution is computed over radial bins (blue curve), with the variation from the 10th to the 90th percentile shown as a shaded area. As is natural for SPH simulations, the highest resolution is obtained where the most particles are: in the disk ($r < 10-30$ kpc). The resolution degrades as we move farther  from the center, to a value $\lambda \sim 30-100$ pc at a radius $R/R_{\rm vir}\simeq 0.1-0.2$, where most of the cold gas is found (see Sect. \ref{overview_cold_gas}), which is approximately the size of a giant molecular cloud. This resolution is comparable to that of other works focusing on the CGM, using different codes and numerical techniques, such as the super-refinement zoom-in simulations of \citet{suresh2019zooming} and the  cosmological box Illustris TNG50 \citep{genel2014introducing, nelson2020resolving}. It is possible that cold gas clumps and filaments in the CGM could be  smaller, implying that our analysis on the origin of cold gas is limited to the scale we are able to resolve. Further convergence studies are needed to investigate the importance of  smaller-scale mechanisms (e.g., Kelvin-Helmholtz and Rayleigh-Taylor instabilities) on gas condensation.

%%%%%%%%%%%

\subsection{Ion abundances and column density maps}

To make a comparison with observations of cold gas around low-redshift galaxies, we computed the ion abundances for two common tracers, HI and Mg II, from our simulation, and their column density distributions.
The HI abundance in the simulation is computed on-the-fly through a primordial chemical network including six species ($e^-$, HI, HII, HeI, HeII, HeIII), accounting for collisional ionization, recombination, and photoionization by the \citet{haardt2012radiative} UV background, with corresponding heating and cooling rates \citep{abel1996modeling}. Gas self-shielding is not taken into account in the simulation.
The abundance of MgII is obtained interpolating from \cloudy~tables, in the same way as in \citet{shen2010enrichment,shen2013circumgalactic}. Tables were obtained by running single-zone models (i.e., thin slabs) for a range of densities and temperatures, assuming photoionization equilibrium under the \citet{haardt2012radiative} uniform UV background (the same background  used for HI and MgII abundance calculations).

%--------------------------------------------------------------------
\subsection{Particle tracking}

\begin{figure}
\centering
\includegraphics[width=0.49\textwidth]{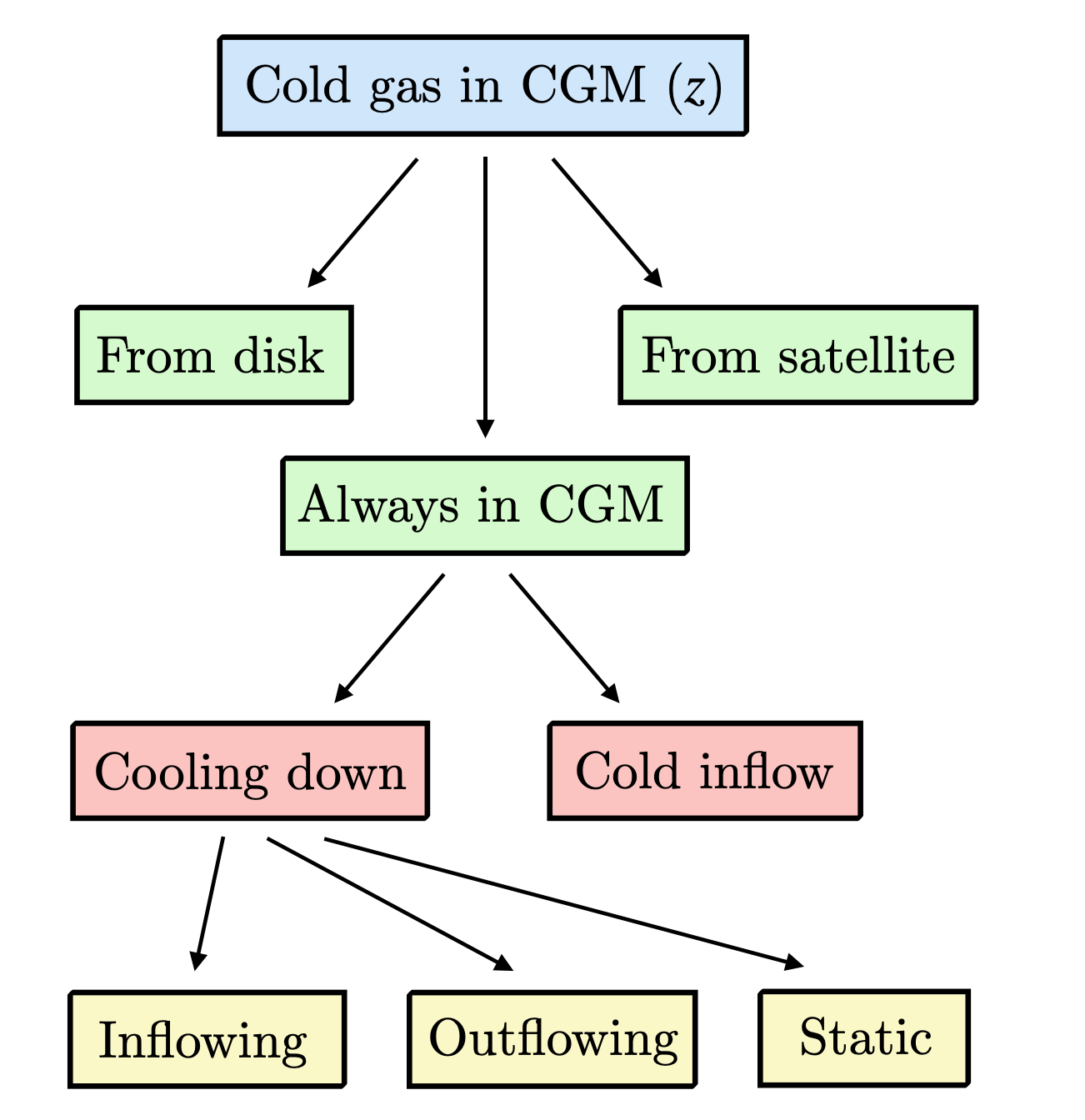}
\caption{After selecting cold gas particles in the CGM (hence excluding gas in the disk of the main galaxy and gas in satellites) at a chosen redshift $z$, their origin was discovered by tracking back in the previous snapshots of the simulation. The green boxes represent location criteria: particles coming from the disk, coming from a satellite, or always present in the CGM. In the red boxes are shown particles  in the CGM that have always been cold (hence in cold inflow since they must have been accreted directly in cold phase at some moment in the galaxy evolution) and particles that cooled down from above the threshold temperature ($T=3\times 10^4$ K). Finally shown are static particles (radial velocity $|v_r|<50$ km s$^{-1}$) and particles with inflowing or outflowing velocity.
     }
\label{diagram_origin}
\end{figure}

SPH particles in the simulations are assigned a unique ID, so that it is possible to track them across snapshots. In this way we are able to check the variation in the  physical properties of parcels of gas (with the mass equal to the mass resolution of the simulation, i.e., around $2\times 10^4\,\msun$) across the evolution of the halo. In particular, we are interested in the temperature evolution to understand how cold gas in the CGM forms, and we follow its trajectory in the halo as well. 

We only analyze   the main halo of the simulation, which is identified by using the Amiga Halo Finder (AHF) \citep{knollmann2009ahf}. Then we find the center of mass of the halo, and we define the disk and CGM in the following way.
For simplicity, we classify as ``disk'' all the gas within $10$ kpc of the center, which is approximately twice the half-mass radius of the stellar component (evolving from $\sim 7$ kpc at $z = 2$ to $\sim 13$ kpc at $z= 0.3$).
The CGM includes all the gas at a distance $10\,{\rm kpc}<r<1.5\rvir$ from the center (i.e., up to a radius of $\sim 110$ kpc at $z=2$ and $\sim 310$ kpc at $z=0.3$).
Throughout the paper, we refer to cold gas as gas with temperature $T<3 \times 10^4$ K (as in \citet{sokolowska2018complementary}) in order to include all neutral gas and gas bearing low ionization potential ions (e.g., HI, MgII, SiII).

To understand the origins of cold gas at a given redshift $z$, each SPH particle that belongs to the cold CGM ($T<3\times 10^4$ K, $r>10$ kpc) at that redshift is tracked back in the previous snapshots until one of the possible origin channels is identified (see Fig. \ref{diagram_origin}). First (green boxes), we identify whether the particle comes from the disk or from a satellite (i.e., belongs to a satellite of the main halo in the  AHF catalog). If not, then the particle could cool down in the CGM from a hot phase or be accreted directly in cold phase (red boxes). Finally, among particles cooling down, we distinguish those with an outflowing velocity (radial velocity $v_r>50\,\kms$) or  inflowing velocity ($v_r<50\,\kms$), or that are static ($-50\,\kms<v_r<50\,\kms$).

Moreover, to understand the trajectory and thermal evolution of gas in the CGM, we track forward gas particles from specific regions of the galaxy. In particular, we consider three setups:
forward tracking of cold disk gas, to understand to what extent it contributes to the cold gas in the CGM, by ejection from the disk;
forward tracking of cold outflowing gas ($T<3\times 10^4$ K, $v_r>50\,\kms$), to understand if this gas survives and reaches the CGM without being heated; and
forward tracking of hot outflowing gas ($T>3\times 10^4$ K, $v_r>50\,\kms$), to understand if cooling in the outflow occurs.

%--------------------------------------------------------------------
%--------------------------------------------------------------------
\section{Results}

%--------------------------------------------------------------------
\subsection{Overview of the cold gas in Eris2k}
\label{overview_cold_gas}

\begin{figure*}
\centering
\includegraphics[width=0.68\textwidth]{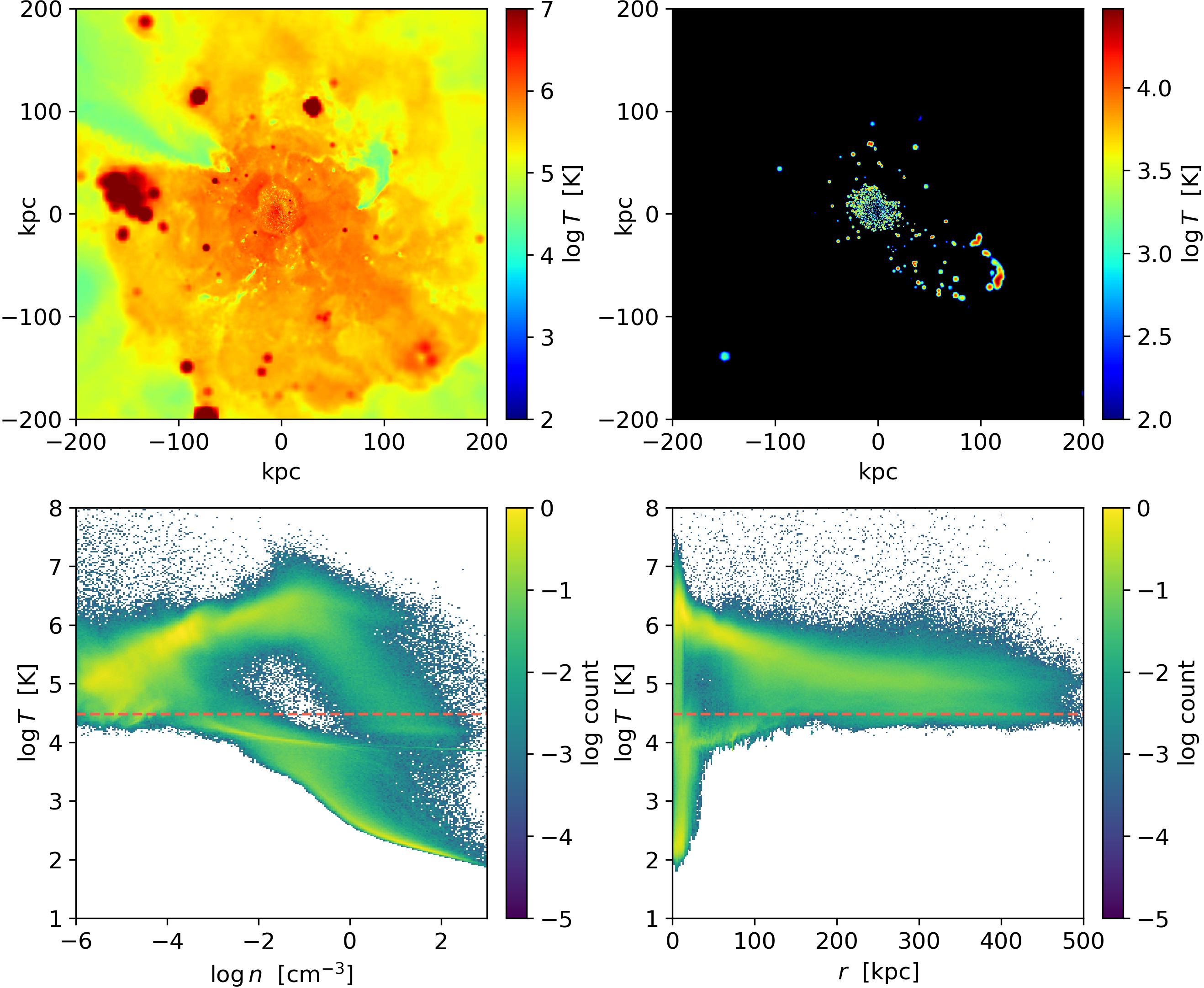}
\caption{
Overview of Eris2k gas temperature at $z\simeq 0.3$. {\bf Upper left}: Temperature map on a slice of  gas that passes through the halo center. Small green patches of gas with temperatures lower than $10^5$ K are visible, ubiquitous, and present especially in the central region within 100 kpc. {\bf Upper right}: Only gas with temperature $T<3\times 10^4$ K is shown from the same slice. Most of the cold gas is located in the central area, and large bubbles are also present  at greater distances. {\bf Lower left}: Distribution of the gas in the whole box in the density-temperature diagram; the red line marks the threshold temperature $T=3\times 10^4$ K. {\bf Lower right}: Distribution in the radial distance--temperature diagram. 
}
\label{overview_panel}
\end{figure*}

\begin{figure*}
\centering
\includegraphics[width=0.80\textwidth]{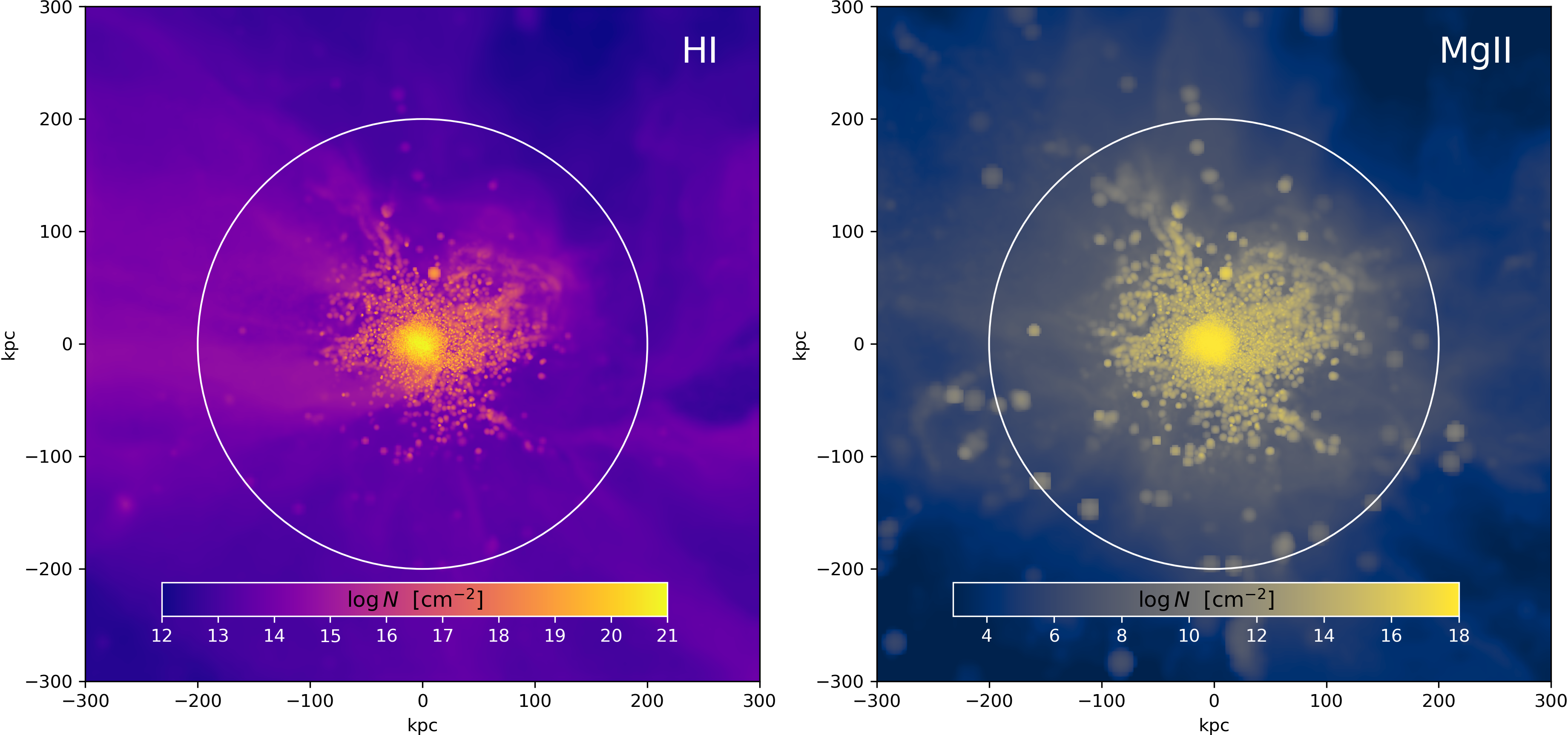}
\caption{
Column densities of HI ({\bf left}) and MgII ({\bf right}) at redshift $z\simeq 0.3$, for a box of size $600^3\,$ kpc$^3$ around the main halo in Eris2k. The white line shows the virial radius of the halo. 
}
\label{HI_MgII_maps}
\end{figure*}

\begin{figure*}
\centering
\includegraphics[width=0.80\textwidth]{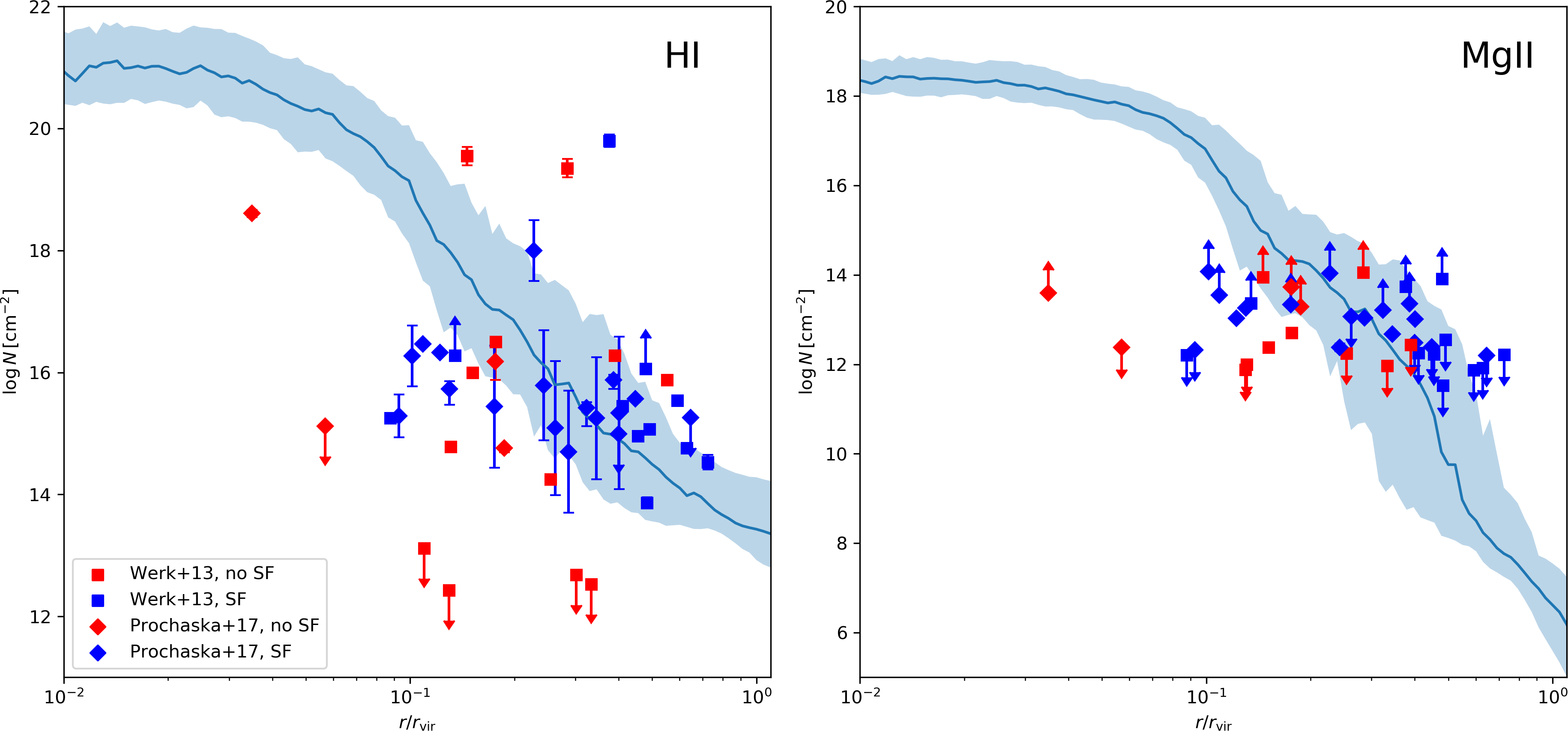}
\caption{
Radial profiles of HI ({\bf left}) and MgII ({\bf right}) column densities, averaging over annuli of thickness 0.5 kpc. The solid line represents the median value, and the shaded region includes the variation from the 10th to the 90th percentile. The observational data from the COS-Halos survey are overplotted, distinguishing between star-forming (blue squares) and non-star-forming (red diamonds) galaxies. The arrows indicate upper and lower limits.
}
\label{N_profile}
\end{figure*}

%% MAPS OF TEMPERATURE
We begin by presenting the main thermal properties of gas in the largest halo in the simulation Eris2k.
A map of gas temperature, for a thin slice passing through the center of the box, is shown in the upper left panel of Fig. \ref{overview_panel}. The upper right panel of the same figure shows only the cold gas (i.e., with temperature lower than $3\times 10^4$ K in our definition). Cold gas is present in the form of numerous small clumps distributed mainly in the region within 100 kpc of  the center. These clumps are not easily seen in the left panel, due to the large amount hot gas in which they are embedded. The temperature can go as low as $10^2$ K in the clumps nearer to the center (i.e., clumps in the disk of the galaxy). The mass of the cold gas in the CGM is $7.6\times 10^{9}\,\msun$, representing  $\sim15.7$\% of the total CGM gas mass; instead, cold gas makes up $\sim 40$\% of the total gas in the disk.

The lower panels of Fig. \ref{overview_panel} show the mass-weighted distribution of gas in the density-temperature plane (lower left panel) and the radius-temperature plane (lower right panel). Cold gas is present with higher mass  in the region at $T\simeq 10^4$ K, corresponding to gas in photoionization equilibrium by the UV background, and  in the tail of dense gas going down to densities $n>1\,\cc$. Gas with high density is favored  to cool down because of shorter cooling times (see Sect. \ref{cooling_gas_cgm}). In the lower right panel of Fig. \ref{overview_panel}, we can see how most of the cold gas is located in the central region, as   was already observed from the temperature slice. Nevertheless, we note that there is a substantial amount of gas below $10^4$ K within a relatively large radius, beyond the disk region, up to $\sim 50$ kpc. Cold gas is also present at larger distances, without showing any specific preference at any radius.

%% ION COLUMN DENSITIES
Figure \ref{HI_MgII_maps} shows the column densities of two main cold gas tracers, HI and MgII, computed for the main halo in Eris2k at  $z\simeq 0.3$. Most of cold gas is clearly seen in the central region of the galaxy ($r<20$ kpc), associated with the disk, with column densities $N_{\rm HI}\simeq 10^{21-22}\,\cm2$ and $N_{\rm MgII}\simeq 10^{17-18}\,\cm2$. Nevertheless, co-spatial clumps of HI and MgII are also seen up to $r\sim 100$ kpc, in the CGM of the galaxy, with sporadic clumps also visible at the virial radius and beyond. In Fig. \ref{N_profile} we show the radial profiles of $N_{\rm HI}$ and $N_{\rm MgII}$, averaging over annuli of radial separations of  0.5 kpc. The profiles are flat in the central region, corresponding to the disk, and decline quickly for $r>0.1\,\rvir$. The $N_{\rm HI}$ profile presents a plateau for $r>\rvir$, which is not seen for the $N_{\rm MgII}$ profile, declining below $10^7\,{\rm cm}^{-2}$ at $\rvir$. This points   to a more extended distribution of HI with respect to MgII, as  also found in other analyses of simulated halos of similar mass \citep{oppenheimer2018multiphase, nelson2020resolving}. Points representing the data from $L^{*}$ galaxies in the COS-Halos survey are shown in the plot \citep{werk2013cos, prochaska2017cos}. The data present a large scatter and a comparison with Eris2k is not straightforward since the observed objects span a wide range of properties, such as different halo masses and star formation rates. Nevertheless, by scaling the impact parameter of the observation with the virial radius, we find many galaxies with column densities compatible to those obtained from the simulation.

%COMPARISON WITH COLD GAS IN OTHER WORKS
 Cold gas extending up to the virial radius is also found   in other works focusing on the thermal properties of the CGM, for simulations run with different codes and different refinement schemes for the CGM gas. \citet{hummels2019impact} pointed out that the abundances of HI and low ionization potential ions increase, at the expense of high ionization potential (as OVI), when the CGM is better resolved.  Cold HI-bearing gas is found in small and dense clumps, that end up mixed with the hot phase in the low-resolution runs. In a similar fashion, \citet{peeples2019figuring} showed that resolving smaller structures in the CGM generally increases the HI column density. \citet{suresh2019zooming} analyzed two zoom-in simulations, finding that the majority of dense ($n>10^{-3}\,\cc$) gas in the galaxy is cold ($T<10^5$ K) up to $0.6-0.7\,R_{\rm vir}$, with a significant fraction extending up to $R_{\rm vir}$. \citet{nelson2020resolving} resolved individual cold clumps by producing HI and MgII maps for massive halos ($M>10^{13}\,\msun$) in the simulation box TNG50, though the CGM is not explicitly targeted in the refinement criteria.

%--------------------------------------------------------------------
\subsection{The origin of cold gas}
\label{origin_cold_gas}

\begin{figure*}
\centering
\includegraphics[width=0.65\textwidth]{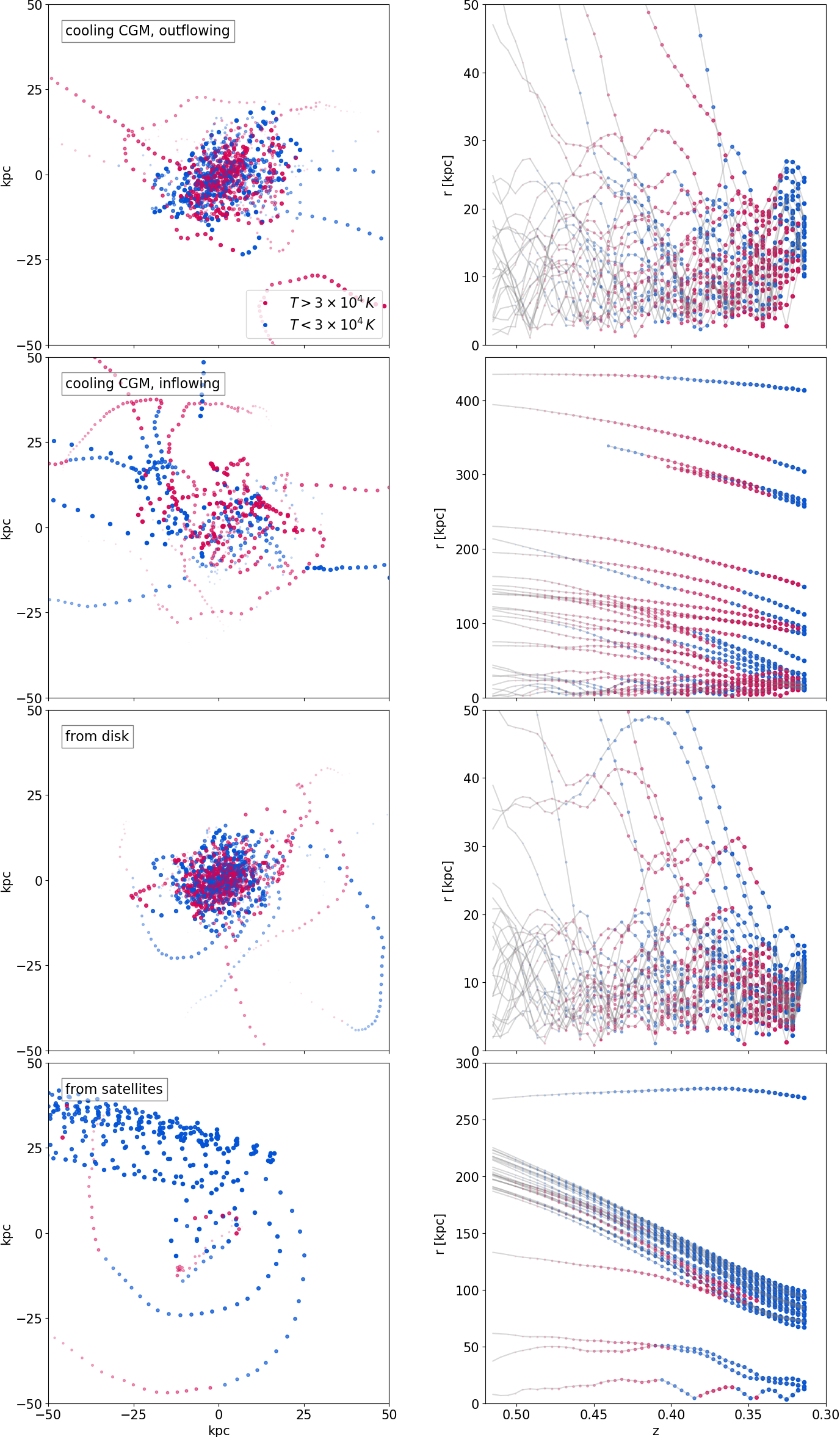}
\caption{
Random selection of 30 SPH particles  chosen from those having the origin ``cooling CGM, outflowing,'' ``cooling CGM, inflowing,'', ``from disk,'' and ``from satellites.'' The backtracking starts from $z\simeq 0.3$ {\bf Left column}: Trajectory of the selected particles in a 2D plane, following them from $z\simeq 0.5$ to $z\simeq 0.3$. The dots become smaller and fainter going back in time, and the color is red  and blue  when the temperature is respectively higher and lower  than $3\times 10^4$. {\bf Right column}: Radial distance from the center of the halo of each particle as a function of redshift.
}
\label{origin_trajectories}
\end{figure*}

\begin{figure*}
\centering
\includegraphics[width=0.95\textwidth]{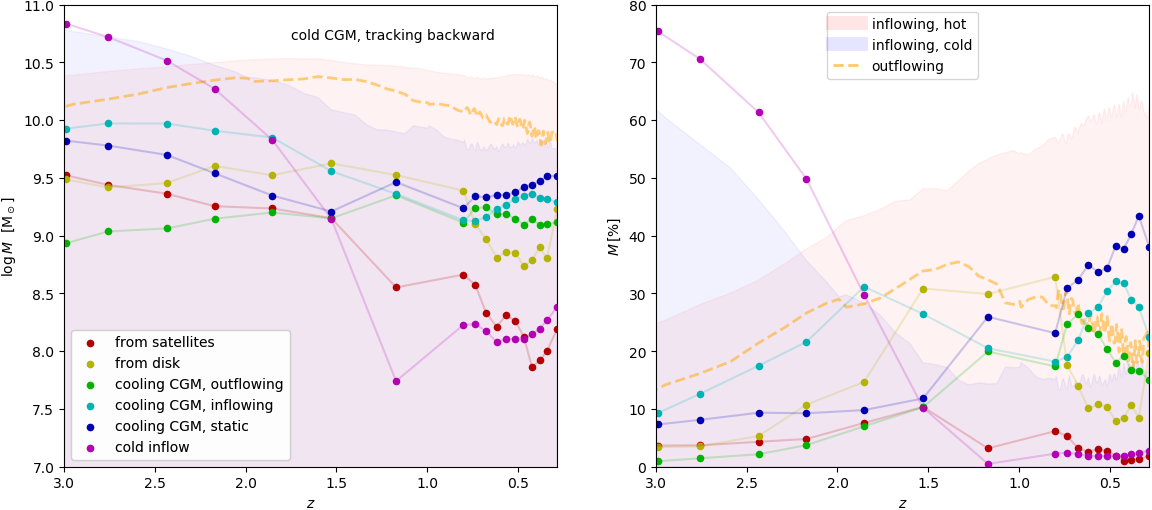}
\caption{
For each snapshot at a given redshift $z$, the gas particles in the cold CGM are tracked back until their origin is identified (according to the sketch in Fig. \ref{diagram_origin}). {\bf Left}: Colored points at a given $z$ representing the total mass of particles having a particular origin, as in the legend. The shaded blue and red areas show the total cold and hot inflowing gas mass, respectively. The orange dashed lines represents the total outflowing gass mass.  
{\bf Right}: Same as the left panel, but the points represent the fraction of gas mass with respect to the initial mass of the cold CGM that is tracked back at a given redshift. In a similar fashion, the inflowing and outflowing gas are represented as a fraction. 
}
\label{origin_tracker}
\end{figure*}

\begin{figure}
    \centering
    \includegraphics[width=0.49\textwidth]{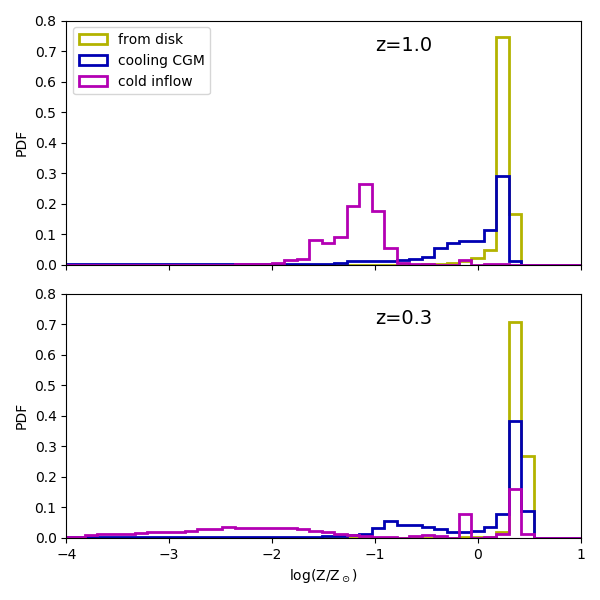}
    \caption{
    Normalized mass-weighted PDF of cold gas particle metallicities for gas classified as having origin ``from disk,'' ``cooling CGM,'' and ``cold inflow'' at redshift $z=1$ (upper panel) and $z=0.3$ (lower panel).  
    }
    \label{zhist}
\end{figure}

\begin{figure}
    \centering
    \includegraphics[width=0.49\textwidth]{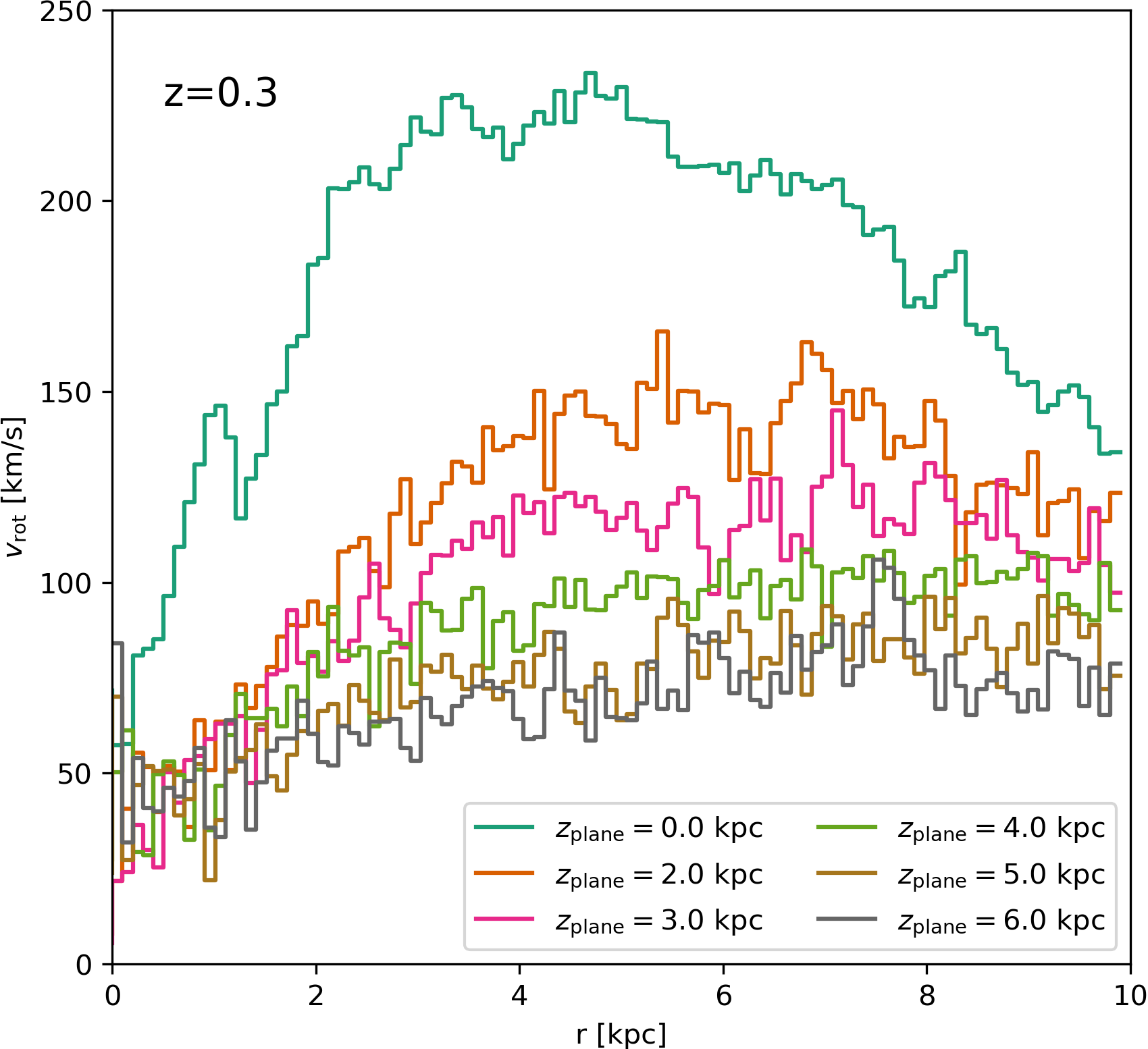}
    \caption{
    Rotation curves in the galaxy disk and in parallel planes at different distances $z_{\rm plane}$. The rotational velocity is computed by defining 100 equal radial bins and averaging particle rotational velocities inside the corresponding annuli.  
    }
    \label{rot_curve_extraplanar}
\end{figure}

Starting from different redshifts, we tracked back the cold SPH gas particles in the CGM to determine one of their possible origins. We identified different possible origins, which are summarized in Fig. \ref{diagram_origin}. First, we adopted a criterion based on the location of the particle to understand if it comes cold from the disk or from a satellite. Otherwise, the particle could have been accreted directly in cold phase or cooled down in the CGM from a hot phase. In the latter case, we also distinguished whether the particle cools down with inward velocity (inflowing, $v_r<-50\,\kms$) or  outward velocity (outflowing, $v_r>50\,\kms$), or is static ($-50\,\kms<v_r<50\,\kms$). 

After applying this algorithm to the cold CGM gas at $z\simeq 0.3$, we found a selection of particles with different origins, and shown in Fig. \ref{origin_trajectories}. In each row, there is  a selection of 30 SPH particles labeled  with the origins ``cooling CGM, outflowing,'' ``cooling CGM, inflowing,'' ``from disk,'' and ``from satellites.'' In the left column we show the trajectories projected in the 2D plane from $z\simeq 0.5$ to $z\simeq 0.3$, with dots becoming progressively smaller and fainter as we go to higher redshifts. Particles are shown in blue when they are cold and in red when they are hot. In the right column the distance of each particle from the center is shown as a function of redshift. We note that the behavior of particles cooling in outflow and coming from the disk is similar: all of them are located very close to the disk (reaching at most a distance $r\sim 25$ kpc from the center), and during their history they are periodically ejected from the disk and then fall back. This means that the only distinction between the two criteria is the moment in the velocity and the distance from the center at which they cool down. We also note that these particles are heated up and cool down many times during their motions. This multiple recycling of gas has also been observed  in \citet{suresh2019zooming}, finding that 60\% of the gas coming from the disk undergoes at least two cycles. A similar behavior is found in the FIRE simulation \citep{angles2017cosmic} for the different types of galaxies analyzed (a massive dwarf spheroidal, Milky-Way mass  systems, and an early-type galaxy): the recycled-to-ejected mass ratio does not show a clear dependence on halo mass and on redshift ($f_\textsc{rec}\simeq 50-95\%$), while the average number of recycles is higher for low-mass systems ($\langle N_\textsc{rec} \rangle \simeq 3-4$ for Milky Way-like galaxies). 

Particles cooling while inflowing show a more regular pattern, with many of them inflowing together as a larger stream, and hence showing a coherent position as a function of redshift $r(z)$, and also cooling down at a similar moment (around $z\simeq 0.4$ for the particles shown). Similarly, also particles coming from satellites have a coherent behavior, being bonded by the gravitational well of the subhalo, and most of them are always cold during their motion. 

Figure \ref{origin_tracker} shows the cumulative result of the origin tracking. From each chosen redshift, we track back the cold CGM gas until we identify the origin. The total mass of gas in each identified origin is shown in the left panel, while we show the percentage with respect to the total cold CGM gas in the right panel. When tracking back from low redshift ($z\simeq 0.3$), most of gas cools down directly in the CGM, with a little contribution from the disk or inflowing streams. The fraction of gas cooling in the CGM decreases at high redshift, making space for gas that is accreted cold from outside the galaxy.

An interpretation for this transition, occurring between $z = 1.5$ and $z=2$, can be found considering the accretion history of the halo. In Fig. \ref{origin_tracker} we show the total hot (red shaded area) and cold (blue shaded area) inflowing gas, and the total outflowing gas (orange dashed line), in solar masses in the left panel and as percentage over the total inflowing plus outflowing gas in the right panel. At $z\simeq 2.2$, when the halo mass is around $3\times 10^{11}\,\msun$, we have the transition between accretion via cold mode and hot mode, as discussed extensively in \citet{kerevs2005galaxies}. Hence, when the galaxy is in the cold mode accretion phase, part of this accreted gas is in the CGM, and explains the gas classified as ``cold inflow''. On the other hand, at lower $z$ most accreted gas is in the hot phase, so the main channel for cold gas in the CGM becomes gas cooling in situ (``cooling CGM'', either ``inflowing'', ``outflowing'', or ``static'' depending on its radial velocity at the moment of the cooling). In addition, the peak of outflow activity is reached between $z=1$ and $z=1.5$, implying that there will be a large contribution of recycled gas to the CGM. At this redshift, we have an increase in ``cooling CGM, outflowing'' and gas coming cold from the disk (i.e., cold gas entrained in outflows). To be more quantitative, the majority of cold gas in the CGM at lower redshift ($z<2$) cooled down in the CGM itself (60-80\%), with a  nonnegligible fraction coming from the disk (10-30\%).  

Gas metallicity can also be used to get insight into the cold CGM gas origin since gas from the disk is expected to be metal enriched, while gas populating the CGM directly from the IGM is expected to be pristine. In Fig. \ref{zhist} we plot the PDF of gas metallicity for cold gas particles that have been classified as originating ``from disk,'' ``cooling CGM'' (without distinguishing   among outflowing, inflowing, and static), and ``cold inflow.'' At both redshifts shown ($z=1$ and $z=0.3$), gas coming from the star-forming disk has solar or super-solar metallicity, while gas coming from cold inflows is mainly metal-free. Only a small fraction of the ``cold inflow'' gas is enriched at $z=0.3$, and we verified that this gas is located (or was recently located) close to the disk at $r=10-15$ kpc, so it is not  classified as ``from disk'' because of  our geometrical definition of disk, but it still displays properties similar to the cold CGM gas coming from the disk. On the other hand, the gas classified as ``cooling CGM'' has a bimodal metallicity, showing that even metal-poor gas is able to cool down from a hot phase in the CGM. It is indeed possible for gas with metallicity as low as $Z=10^{-3}$ to cool down from temperatures $T=10^{5-6}$ K when the density is high enough ($n>10^{-2}\,\cc$), as explored in more detail in Sect. \ref{cooling_gas_cgm}. We note that the  Eris2k simulation  includes metal diffusion due to turbulent mixing. However, as shown in \citep{shen2010enrichment} (e.g., their Figure 9), this model only acts on a subgrid scale, and does not appear to significantly alter large-scale mixing between pristine gas in accretion streams and gas outflowing from the disk.

To make a comparison with previous works on the origin of CGM gas, we have to consider that  gas is usually tracked back without stopping the tracking when the particle is heated; in other words,  the studies focus on understanding where particles come from, regardless of when it cools down. Tracking back farther in our simulation the particles labeled as ``cooling CGM, static'' from $z\simeq 0.3$, we find that $\sim 55\%$ come  from the disk (even if cooling down later in the CGM) and $\sim 33\%$ come  from $r>\rvir$. In this way, we obtain that $\sim 65\%$ of cold CGM gas at  $z\simeq 0.3$ comes from the disk, and $\sim 30 \%$ is accreted from the IGM. The importance of galactic winds in feeding the CGM has already been assessed in previous works \citep{sokolowska2018complementary}, with a contribution spanning from 30\% \citep{ford2014tracing, hafen2019origins} to 70\% 
\citep{suresh2019zooming}. Nevertheless, the relative importance among different origins (IGM accretion, outflows, and satellites) appears to be strongly dependent on the halo mass and the redshift from which the gas is tracked back. 

The contribution of different components to the cold CGM gas also reflects  its dynamical properties. In particular, observational evidence of the contribution of galactic fountains to the cold CGM gas budget lies in the measured vertical gradient of the rotational velocity of gas in planes above and below the mid-plane of the galaxy (known as extra-planar gas). This has been observed in nearby spiral galaxies \citep[e.g.,][]{boomsma2004high, fraternali2004extra, oosterloo2007cold, voigtlander2013kinematics, li2021kinematic}, and  is interpreted as a loss of angular momentum when gas ejected from the disk interacts with the hot CGM medium. In Fig. \ref{rot_curve_extraplanar} we show rotation curves for the Eris2k disk ($z_{\rm plane}=0$) and parallel planes at different distances from the disk ($z_{\rm plane}=2-6$ kpc). Each plane is actually a slab of thickness 0.4 kpc, and the rotational velocity $v_{\rm rot}$ is computed by averaging over 100 equal radial bins spanning the range $0<r<10$ kpc. The gas retains a high rotational velocity up to 6 kpc, with peak velocities of 50-100 km/s. The rotational velocity globally decreases as we take planes that are more distant from the disk plane, with velocity gradients between -10 and -20 ${{\rm km}\,{\rm s}^{-1}\,{\rm kpc}^{-1}}$, depending on the radius at which we measure $v_{\rm rot}$. These values are in good agreement with the above-mentioned observations of extra-planar gas and other theoretical studies \citep{melioli2009hydrodynamical, marinacci2011galactic, armillotta2016efficiency}.

A more detailed exploration of the disk-corona interaction is needed, and in principle it can be pursued by tracking SPH gas particles in the two components, following the evolution of their angular momentum. In a similar way, we suggest that the relative contribution of hot and cold mode accretion could lead to a correlation of the angular momentum of accreting gas with the angular momentum of the CGM. We defer to a future work for a detailed analysis of CMG dynamical properties.

%--------------------------------------------------------------------
\subsection{Forward tracking of disk gas}

\begin{figure*}
\centering
\includegraphics[width=0.95\textwidth]{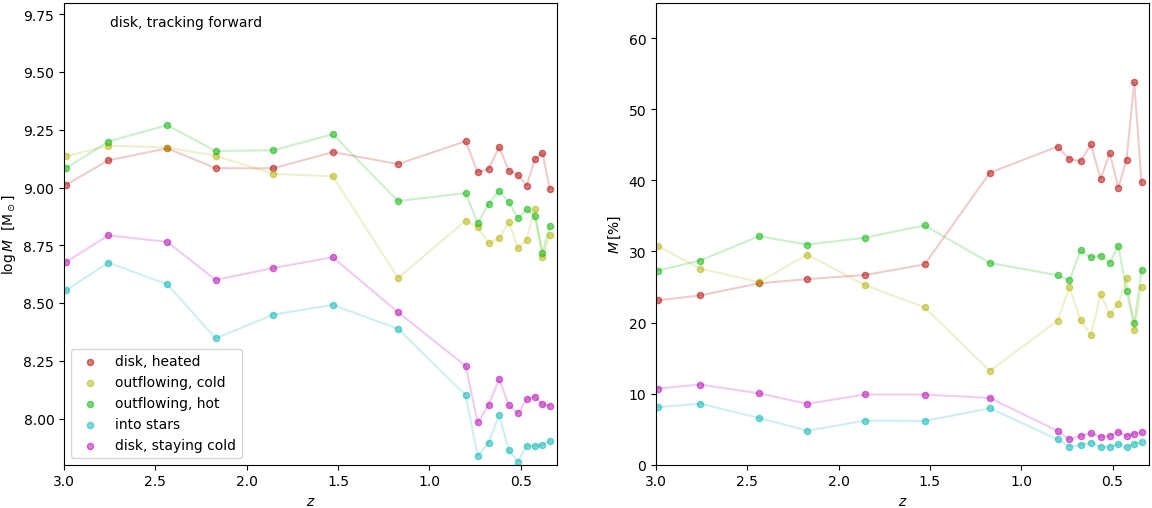}
\caption{
For each snapshot at a given redshift $z$, gas particles that  are cold and in the disk are tracked forward until an outcome is identified: the particle is heated within the disk, the particle is launched in an outflow but is still cold, the particle is launched in an outflow and is heated above $T_0$, or the particle is converted into stars. If none of the previous outcomes is found earlier than $z=0.3$, it means that the particle remains cold in the disk. {\bf Left}: Total mass of the gas in each outcome. {\bf Right}: Fraction of the gas mass in each outcome, with respect to the initial cold disk gas at the redshift $z$ at which the forward tracking is started.  
}
\label{disk_fate_tracker}
\end{figure*}
Starting from a range of different redshifts, we track forward the cold gas in the disk to understand its fate. The forward tracking is stopped when the particle is heated, when the particle is ejected in an outflow (i.e., radial velocity $v>50$ km/s, distinguishing between hot and cold outflowing gas), or when the gas is converted into a star particle. Particles that have not undergone any of these outcomes at $z=0.3$ are labeled as ``staying cold'' in the disk (with only a negligible percentage going out from the disk with low velocities $v<50\,\kms$). The result of this analysis is shown in Fig. \ref{disk_fate_tracker};   the total mass in each outcome is in the left panel, and the fraction with respect to the initial cold disk gas (at the redshift we start the tracking) is in the right panel. Starting the tracking from any redshift $z>1.2$ gives approximately the same result: about 30\% of the gas is outflowing in hot phase, $\sim$30\% is outflowing in cold phase, and $\sim$25\% is heated but stays in the disk; a smaller fraction ($\sim$10\%) remains cold and inside the disk, and a similar amount fuels star formation. A transition occurs at $z\simeq 1.5$, where the amount of hot gas remaining in the disk increases to $\sim 45\%$, and the outflowing gas decreases (both hot and cold) to 15-20\%, while  gas converted into stars also slowly decreases to a few percent. The reduction of gas in outflow  was already highlighted in Fig. \ref{origin_tracker} (orange dashed line), and it is  evident here how it is associated with a decrease in star formation activity. As a consequence, there is also more heated gas that is not able to escape the galaxy potential well. 

%This points to a high star formation activity (since gas is indeed heated by the feedback), whereas the gas is not able to escape the potential well of the galaxy, hence a reduction of the outflow activity as well (dashed line in Fig. \ref{origin_tracker}).

%--------------------------------------------------------------------
\subsection{Forward tracking of outflowing gas}

\begin{figure}
\centering
\includegraphics[width=0.49\textwidth]{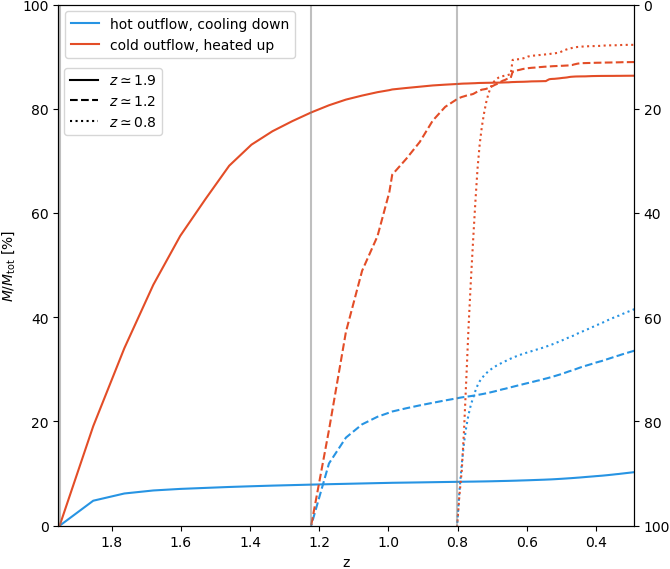}
\caption{
Forward tracking of outflowing gas (defined as gas with outward velocity $v>50$ km s$^{-1}$), starting from redshift $z\simeq 1.2$ (solid line), $z\simeq 1.2$ (dashed line), and $z\simeq 0.8$ (dotted line). The red line shows the cumulative fraction of cold outflowing gas that is heated, while the blue line represents the hot outflowing gas that cools down.
}
\label{outflow_forward_tracker_mass}
\end{figure}

\begin{figure*}
\centering
\includegraphics[width=0.95\textwidth]{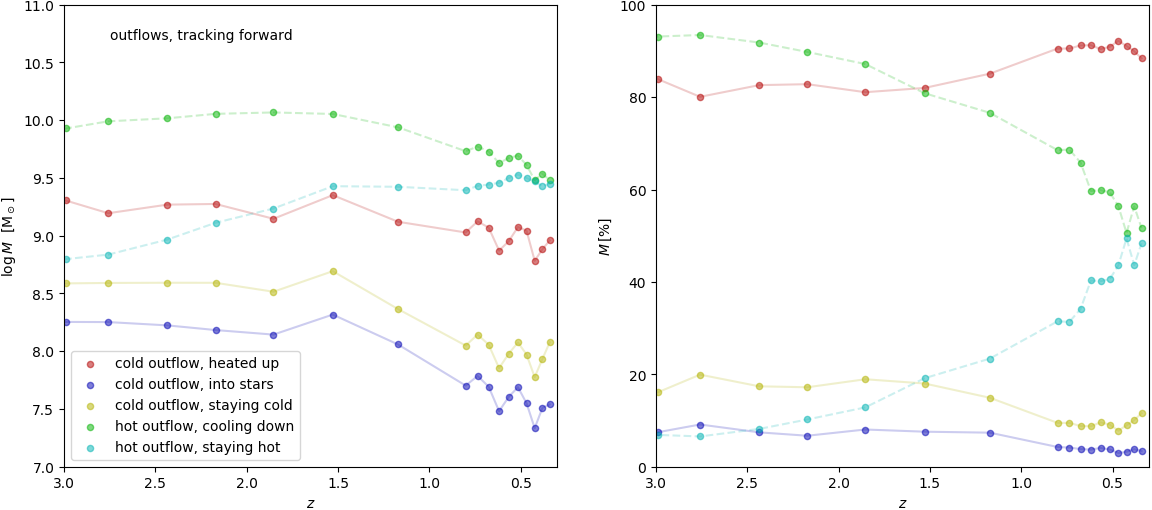}
\caption{
For each snapshot at a given redshift $z$, outflowing gas particles are tracked forward in time. For cold gas particles (solid lines), the tracking is stopped when they are heated above $T=3\times 10^4$ K (``heated up'') or   form stars (``into stars''), while particles not undergoing either   outcome  at $z=0.3$ are labeled   ``staying cold.'' Similarly, the tracking of hot particles (dashed lines) is stopped if they cool down or form stars, so that the possible outcomes are ``cooling down,'' ``into stars,'' or otherwise ``staying hot'' (although a hot gas particle needs to become cold before being eligible for star formation).
{\bf Left}: Total mass of the gas in each outcome. {\bf Right}: Fraction of the gas mass in each outcome, with respect to the initial cold outflow mass for heated up, into stars, and staying cold outcomes, and with respect to the initial hot outflow mass for cooling down and staying hot outcomes.
}
\label{outflow_fate_tracker}
\end{figure*}

\begin{figure*}
\centering
\includegraphics[width=0.95\textwidth]{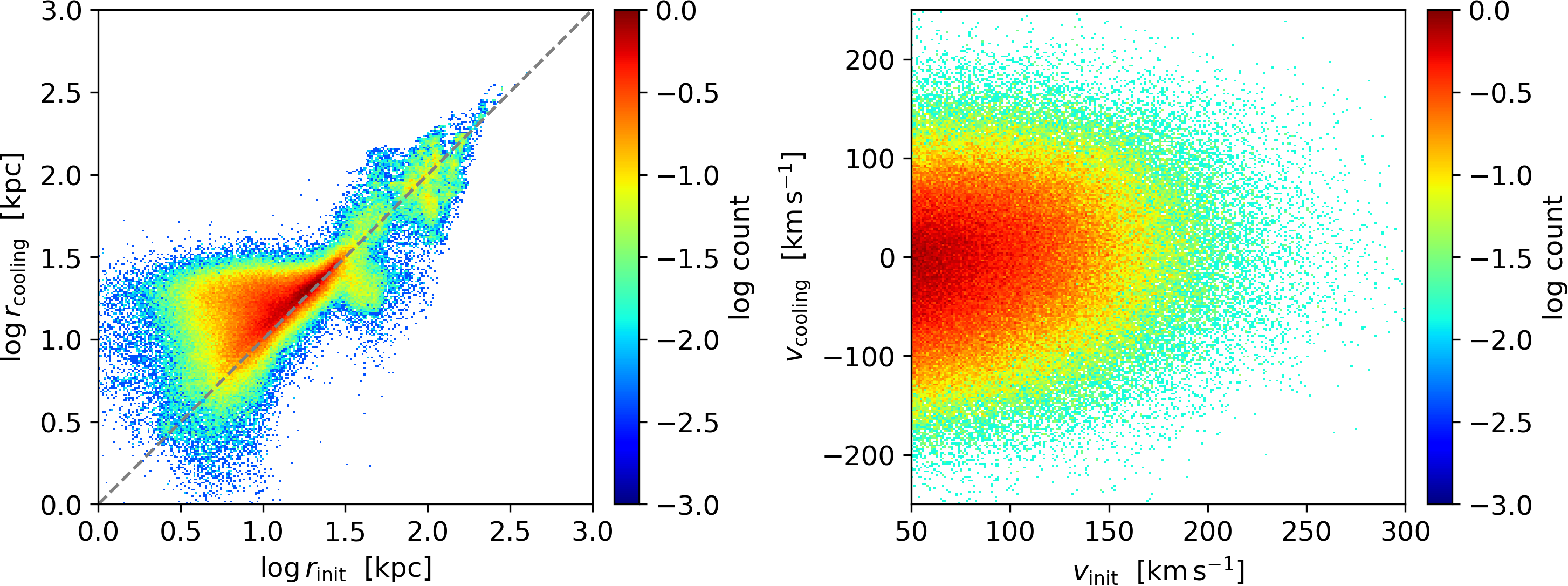}
\caption{
Hot outflowing gas is tracked forward from $z\simeq 0.5$ until it cools down. The two panels show the mass-weighted particle distribution according to their initial radius ($r_{\rm init}$) and the radius at which they cool down ($r_{\rm cooling}$) in the {\bf left} panel, and according to their initial radial velocity ($v_{\rm init}$) and the velocity at which they cool down ($v_{\rm cooling}$) in the {\bf right} panel. The two panels show the distribution of particles, weighted by their mass, according to their initial radius (x-axis), and the radius at which they cool down (y-axis). 
}
\label{hot_outflow_point}
\end{figure*}

\begin{figure}
\centering
\includegraphics[width=0.49\textwidth]{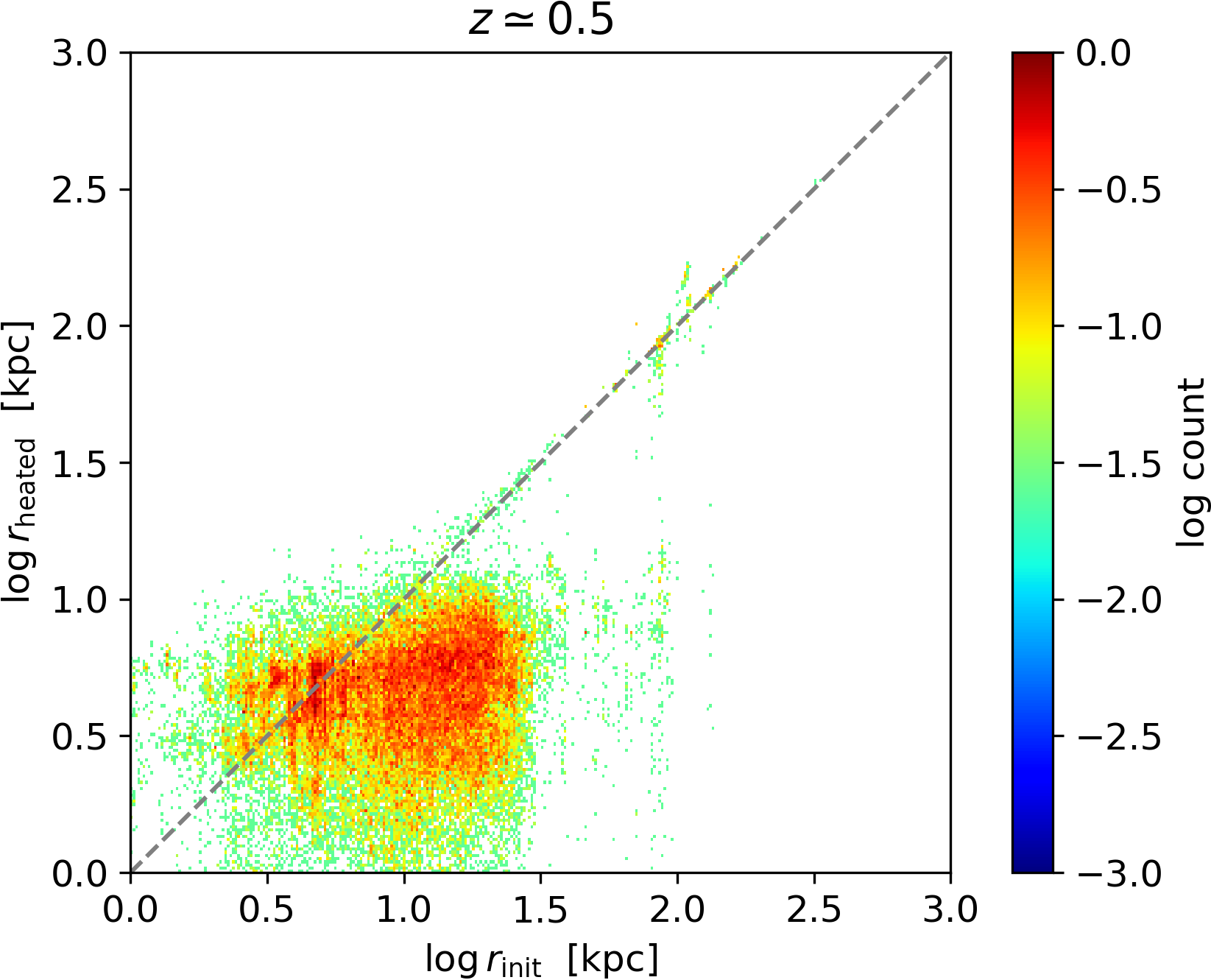}
\caption{
Cold outflowing gas is tracked forward from $z\simeq 0.5$ until it is heated up. The plot shows the distribution of particles, weighted by their mass, according to their initial radius at $z\simeq 0.5$ (x-axis) and the radius at which they are heated (y-axis).
}
\label{cold_outflow_point}
\end{figure}

Finally, we track  forward outflowing particles, in the hot and cold phase, separately. In the first case, the tracking is stopped when a particle cools down ($T<3\times 10^4$ K), while conversely we stop tracking a cold particle  when it is heated up ($T>3\times 10^4$ K) or converted to a star particle. Figure \ref{outflow_forward_tracker_mass} shows the evolution of cold gas heated up (red line) and hot gas cooling down (blue line) starting the tracking from three different redshifts ($z\simeq 2$, 1.2, and 0.8). In all  three cases, around 90\% of the cold gas is heated up in the outflow, allowing 10\% of gas to survive in cold phase. The  timescale for cold gas survival (estimated as the time at which 50\% of gas is heated) is around 300-400 Myr in all cases. On the other hand, the amount of gas cooling down in hot outflows depends strongly on the initial redshift at which the outflow is launched. In particular, only a  small fraction ($\sim 10$\%) of outflows launched at redshift $z\simeq 2$ cool down; this fraction increases to 40\% when the outflow is launched at $z\le 1.2$. Checking the average properties of hot outflowing gas, we find that the average distance from the center changes from $\langle r \rangle_{z\simeq 2} \simeq 150$ kpc to $\langle r \rangle_{z\simeq 0.8} \simeq 75$ kpc (with a corresponding drop in average velocity from $\langle v \rangle_{z\simeq 2} \simeq 100\,\kms$ to $\langle v \rangle_{z\simeq 0.8} \simeq 80\,\kms$). Because it is  closer to the central region, gas that is  launched later is  more dense, and it  cannot reach greater distances due to lower velocity; as a consequence, the cooling time is smaller, and this gas is favored to cool down.

A cumulative analysis of the fate of the cold and hot phases of outflowing gas is shown in Fig. \ref{outflow_fate_tracker}. The solid lines show the cold gas in outflow heated up, remaining cold, and converted into stars. In the right panel we can see that less than 20\% of the cold gas can survive in cold phase, in agreement with what was already observed in Fig. \ref{outflow_forward_tracker_mass}, while 5-10\% of cold gas is forming stars. On the other hand, the dashed lines show the hot gas cooling down and the gas remaining hot. In this case, we see that only $\sim$10\% of gas cools down at high redshift, while this fraction increases to almost 50\% for $z<0.75$. As noted in the previous paragraph, this is due to the outflowing gas being launched from a region closer to the center, which is denser and hence favors gas cooling. Nevertheless, we note that radiation feedback from stellar sources is not included in this simulation, and some studies have shown the importance of this mechanism in heating the cold gas in outflows \citep{decataldo2017molecular}. Hence, our study could overestimate the amount of the cold gas surviving in outflow, and the amount of gas cooling down from the hot phase. 

To inspect further the dynamics of cooling gas in outflow, we analyzed the position and velocity of the tracked gas at the moment of cooling (Fig. \ref{hot_outflow_point}). The left panel shows the distribution of hot outflowing SPH particles according to the radius at the beginning of the tracking (i.e., initial distance from the center) and the radius at which it cools down. Analogously, the right plot shows the initial velocity and the velocity at the moment of cooling. The particles represented here are tracked forward starting from $z= 0.5$, so $r_{\rm init}$ and $v_{\rm init}$ are not necessarily the position and velocity at the moment of the launch of the outflow;  the particle could  already in outflow since a previous snapshot at $z>0.5$. $r_{\rm init}$ spans a range from 3 kpc to $\sim$100 kpc, and most particles cool down immediately or after travelling at most $\sim$ 30 kpc. For some particles we have $r_{\rm cooling}<r_{\rm init}$, meaning that they have reached the turnover point and they are falling back toward the center of the halo. This is also clear from the velocity distribution: most particles cool down at at $v_{\rm cooling} \simeq 0$ (i.e., at the turning point) and the distribution is almost symmetrical between particles cooling down while still outflowing or falling back to the center. This configuration points to a galactic fountain, which has already been shown to be one of the main contributors to the total CGM cold gas budget in previous works \citep{suresh2019zooming}. We also note that particles with higher $r_{\rm init}$ travel only a short distance before cooling because   their velocity is in general lower, either because they are far from the central region where stellar feedback is more effective, or they are already close to the turnover point of an outflow launched at an earlier time.

Similarly, in Fig. \ref{cold_outflow_point} we plot the distribution of cold outflowing particles relative to the radius at the beginning of the tracking and to the radius at which they are heated. 
In this case most particles start their journey at a distance $r<30$ kpc, where most of cold CGM gas is located (see also Fig. \ref{overview_panel}). These particles are then heated after the turnover at radius $r_{\rm heated}<10$ kpc (i.e., within the disk), suggesting stellar feedback as the main heating mechanism.

%--------------------------------------------------------------------
\subsection{The cooling of gas in the CGM}
\label{cooling_gas_cgm}

\begin{figure*}
\centering
\includegraphics[width=0.98\textwidth]{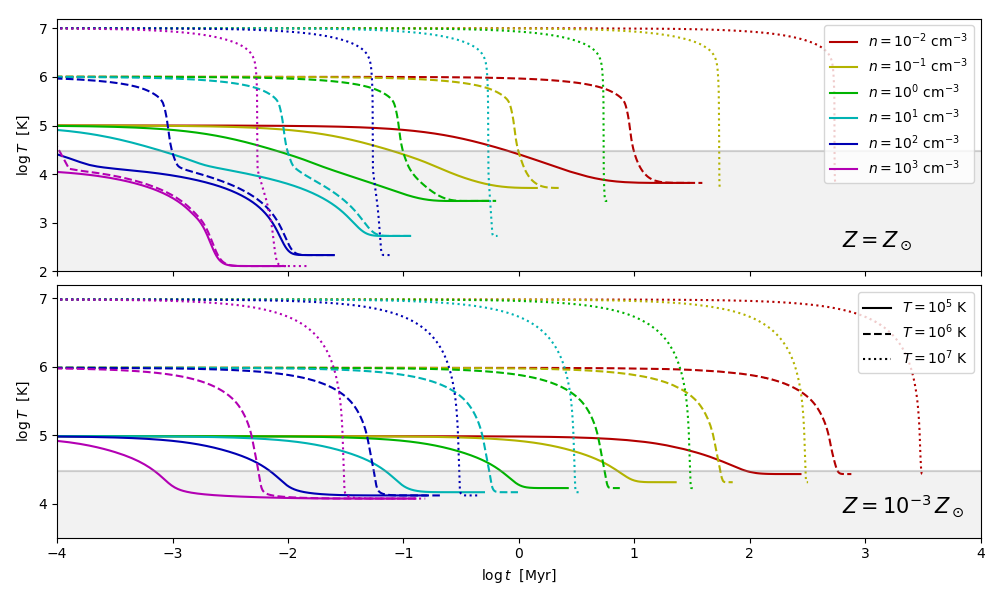}
\caption{
Temperature evolution of gas with different densities (from $n=10^{-2}$ cm$^{-3}$ to $n=10^{3}$ cm$^{-3}$), starting from different initial temperatures ($T=10^5$, $10^6$, and $10^7$ K). The gas has metallicity $Z=Z_\odot$ in the upper panel, and $Z=10^{-3}\,Z_\odot$ in the lower panel. The gray shaded area marks the temperature $T=3\times 10^4$ K, which in our work is defined as the threshold temperature between cold and hot gas. 
}
\label{temperature_evolution}
\end{figure*}

\begin{figure*}
\centering
\includegraphics[width=0.41\textwidth]{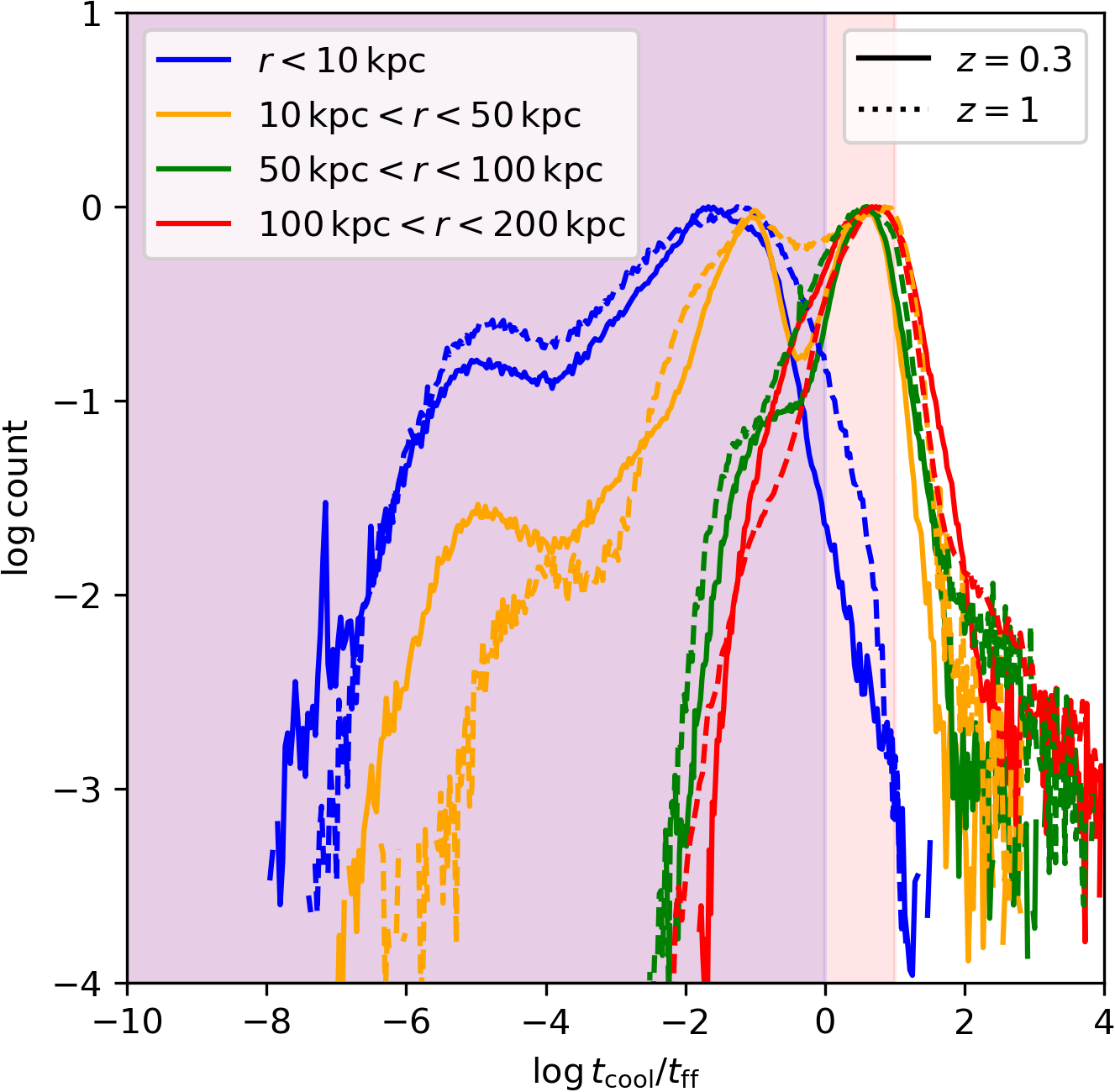}
\hspace{0.5cm}
\includegraphics[width=0.49\textwidth]{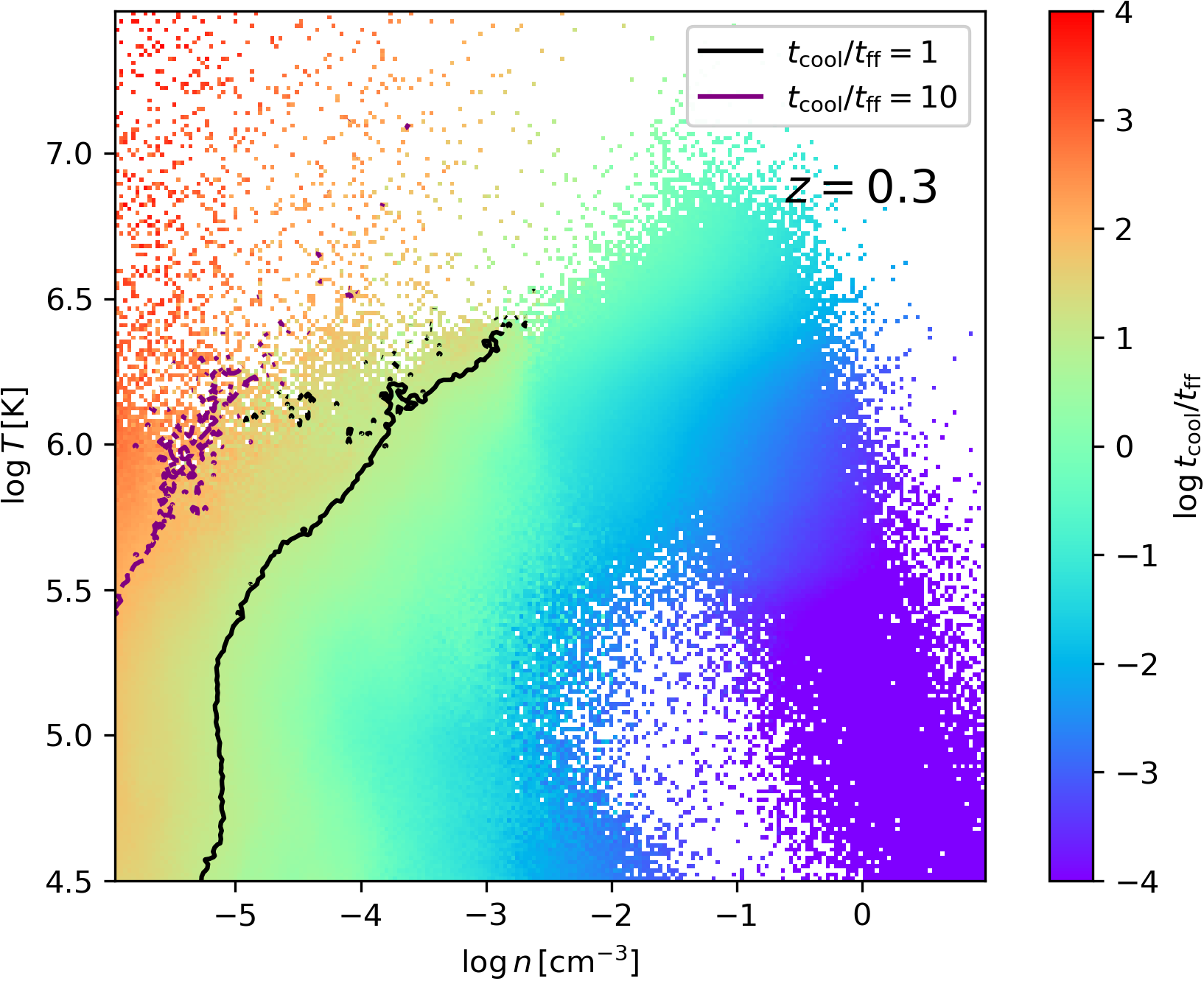}
\caption{
Properties of the hot ($T>3\times 10^4$ K) CGM gas according to its $\tcool/\tff$ ratio.
{\bf Left}: Distribution of $\tcool/\tff$ at two different redshifts, $z=0.3$ (solid line) and $z=1$ (dashed line).  Different colors indicate gas belonging to different radial distance bins, while the shaded regions highlight the $\tcool/\tff<1$ criterion for precipitation-regulated feedback, and the less stringent $\tcool/\tff<10$ condition.
{\bf Right}: Distribution of hot CGM gas at $z=0.3$ in the $n$-$T$ phase diagram, color-coded by the average value of $\tcool/\tff$. The contour lines show the $\tcool/\tff=1$ and $10$ boundaries.
}
\label{tratio_hist}
\end{figure*}

In our analysis of the origin of cold gas in the CGM (Sect. \ref{origin_cold_gas}), we  found that an  important contribution, especially at $z<2$ is from gas cooling down from higher temperatures. This means that detachment from the hot phase and the cooling down is a process that can happen at CGM densities and on timescales comparable to the dynamic timescales of the galaxy. In order to asses this point, we compute the temperature as a function of time for gas with different densities and initial temperatures, adopting the same cooling function used in the Eris2k simulation and assuming constant density during the cooling. The results are shown in Fig. \ref{temperature_evolution}, for gas at solar metallicity (upper panel) and $10^{-3}\,Z_\odot$. The different colors stand for the different gas densities, from $n=10^{-3}$ to $10^3\,\cc$, while the line styles represent the different initial temperatures $T=10^5$, $10^6$, and $10^7$ K. Basically, all gas with density $n \ge 10^{-2}\,\cc$ (which is the average density for gas at a distance $r=10-30$ kpc at $z\simeq 0.3$) can cool down in less than 50 Myr at $Z=Z_\odot$, and in less than 500 Myr for the low-$Z$ case, which allows the gas in the CGM to cool down spontaneously on a short timescale   compared to the galaxy evolution. In addition, gas in outflows   typically has higher densities ($1-100\,\cc$ in the simulation), and consequently with cooling timescales on the order of 10 Myr, even at the highest initial temperatures ($T\sim 10^7$ K).

The condensation of cold clumps in the warm-hot intergalactic medium (WHIM) was studied analytically by \citet{binney2009high} (who had doubts about the efficiency of this process); they  showed via linear perturbation analysis that thermal instabilities in a hot stratified corona are suppressed by buoyancy and thermal conduction. Nevertheless, this mechanism regained favor thanks to studies taking into account the rotation of the corona \citep{nipoti2010thermal, sormani2019effect, sobacchi2019effect} and studies that explored the nonlinear nature of this process \citep{mccourt2012thermal, sharma2012thermal}. In the latter, the possibility of gas cooling and fragmentation within the CGM appears to be related to the ratio $\tcool/\tff$, where $\tff$ is the free-fall time into the potential well of the galaxy. In particular, simulations in these studies find a value $\tcool/\tff<1-10$ to be a necessary condition for condensation. The precipitation-regulated feedback model \citep{sharma2012thermal, voit2017global} manages to explain why galaxies self-regulate to this value of $\tcool/\tff$: outflows promote condensation of adiabatically uplifted gas, which eventually falls back and fuels star formation (and eventually stellar feedback), until the outflows become stronger and the gas is heated too much to allow condensation. However, we mention that \citet{saeedzadeh2023cool} highlighted how the $\tcool/\tff$ criterion does not always capture the susceptibility of gas to condensation:    high-$\tcool/\tff$ gas still cools down in the presence of large-scale perturbations \citep{choudhury2019multiphase} and low-$\tcool/\tff$ gas is unable to condense when the ratio decreases only for a short period of time (e.g., when $\tff$ increases suddenly in the wake of a satellite).

We explore the thermal instability of CGM gas in Eris2k by plotting histograms of $\tcool/\tff$ for hot ($T>3\times 10^4$ K) gas located at different radial distances from the halo center at $z=0.3$ and $z=1$ (left panel of Fig. \ref{tratio_hist}). Most gas in the CGM (hence $r>10$ kpc, which is the typical disk radius assumed in this work) up to 50 kpc has $\tcool/\tff \ll 1$, making the precipitation scenario a feasible channel for the origin of cold gas. This region close to the disk is where most of cold gas clumps are located (see Fig. \ref{overview_panel}), and where the distribution of cooling gas in outflow peaks (see red lines in the left panel of Fig. \ref{hot_outflow_point}). Gas at greater distances ($r>50$ kpc) show a $\tcool/\tff$ distribution peaking around $\sim 5$, but   still with  a large fraction of gas with $\tcool/\tff<1$, especially at higher $z$. In general, we observe that the amount of unstable gas ($\tcool/\tff<1$)  at high radii increases with redshift. 
To better show the physical conditions of unstable gas, in the right panel of Fig. \ref{tratio_hist}, we show the 2D $n$-$T$ phase diagram of the hot CGM ($r>10$ kpc) at $z=0.3$, where the gas in each bin has been color-coded according to the average value of $\tcool/\tff$. As expected already from the analysis in Fig. \ref{temperature_evolution}, high-density gas is unstable because of the short cooling timescale. CGM gas at lower densities ($n<10^{-3}\,\cc$) might still be unstable gas when the temperature is not too high ($T<10^6$ K).

We note that the value of the ratio $\tcool/\tff$ for the gas depends on the maximum resolution reached in the simulation since the cooling time is shorter when high-density gas is resolved ($\tcool \sim n^{-1}$). Hence, at higher resolution it is possible that  the distribution of $\tcool/\tff$ in Fig. \ref{tratio_hist} would move farther to the left, until convergence, implying even more thermally unstable gas. 

%--------------------------------------------------------------------
%------------------------------------------------------------

\section{Conclusions}

In the present work we   analyzed the history of the gas in the main halo of the simulation Eris2k, by tracking SPH particles properties through the evolution of the halo. In particular, we focused on cold gas ($T<3\times 10^4$ K) in the CGM ($10\,{\rm kpc} < r < 1.5\,\rvir$), making up around $15\%$ of the total CGM gas mass at redshisft $z\simeq 0.3$, concentrated mainly in small clumps in the region within 50 kpc.

We tracked back particles from cold gas in the CGM, starting from different evolutionary stages of the halo. At high redshift ($z>2.5$), the CGM is dominated by gas that flows into the halo via cold accretion streams (cold mode accretion), so that the majority of cold CGM gas is accreted directly from the IGM. At lower redshifts ($z<2$), the halo is mainly accreating via the hot mode, and the CGM is dominated by shock-heated and recycled gas. In this regime, we show that most cold CGM gas has cooled down from the hot phase ($\sim$ 40\% of the cold CGM gas is approximately static at the moment of cooling, up to $30\%$ is in inflow). Outflows also contribute significantly to the cold gas budget, especially around $z\simeq 1.5$, which is the peak of outflow activity; cold CGM gas is indeed enriched by gas cooling within the outflow (10-20\%), and cold gas lifted from the disk (up to $30\%$ at $z\simeq 1.5$). We also find that gas can undergo multiple cooling and heating phases, especially particles that are recycled by falling back into the disk after being in outflow.

To inspect further the thermal evolution of the gas in the galaxy, we tracked separately particles in the different components of the galaxy:
\begin{itemize}
    \item Disk: For $z<1.2$ most of gas is expelled in outflows both in cold and hot form ($\sim 30\%$ each), while later more than $60\%$ of the gas remains in the disk, even when heated, due to high star formation and a decreased outflow activity;
    \item Cold outflows: Starting the tracking from different $z$, we always find that only $\sim 10\%$ of the gas is able to survive in cold form until $z\sim 0.3$;
    \item Hot outflow: a higher
fraction of gas launched at lower $z$ tends to cool down  ($\sim 40\%$ from $z\simeq 0.8$) since the central region is more compact and the outflowing gas is initially denser.
\end{itemize} 
We note that the amount of cold gas in outflows could be overestimated since the simulation does not include radiative transfer on the fly, and cold clumps are subject to the intense radiation field of stars in the disk.  

We obtained an insight into the scenario of gas cooling directly in the CGM by computing the cooling time of gas at different initial temperatures and density values. Gas with density $n\geq 10^{-2}\,\cc$ is able to cool down in less than 50 Myr, which is compatible with the spontaneous cooling that we observe in the simulation. In addition, gas at a distance $r<50$ kpc (where most of the cold CGM gas is located) satisfies the condition $\tcool/\tff<1-10$, which allows for gas fragmentation and cooling in the CGM (precipitation-regulated feedback model).

Our analysis was performed on a zoom-in simulation, which allowed for the high spatial resolution (maximum resolution of $\sim 5$ pc in the CGM) needed to resolve the cold phase, which would otherwise be mixed with the hot phase. The drawback of this approach is that it allowed us to study only one sample, specifically a Milky Way-mass galaxy. We expect some differences in systems with different masses, as pointed out by similar analysis on large cosmological boxes \citep{hafen2019origins, nelson2020resolving}; for example, the contribution of winds to the CGM gas budget is seen to decrease with halo mass, while the amount of gas stripped from satellites increases \citep{hafen2019origins}.
Hence, to improve the current analysis on the origin of cold gas, the following steps are needed: First,   galaxies  with different properties should be analyzed (mass, star formation rate), and second, the effect of resolution on the amount of cold gas should be investigated  by comparison with simulations on smaller scales.

\begin{acknowledgements}

The simulations were performed using the resources from the National Infrastructure for High Performance Computing and Data Storage in Norway, UNINETT Sigma2, allocated to Project NN9477K. We acknowledge the support from the Research Council of Norway through NFR Young Research Talents Grant 276043. We also acknowledge use of the Python programming language, Astropy \citep{AstropyCollaboration2013}, Matplotlib \citep{Hunter2007}, NumPy \citep{VanderWalt2011},
Pynbody \citep{pontzen2013pynbody}. We thank the anonymous referee for the insightful comments and suggestions, which significantly helped us improve this work.

\end{acknowledgements}

\bibliographystyle{aa}
\bibliography{biblio}

\end{document}